\documentclass[%
reprint,
amsmath,amssymb,
aps,
pra,
]{revtex4-2}

\usepackage{graphicx}
\usepackage{dcolumn}
\usepackage{bm}
\usepackage{hyperref}
\usepackage[mathlines]{lineno}
\usepackage{bm}
\usepackage{float}
\usepackage{cleveref}
\usepackage{wasysym}
\usepackage[makeroom]{cancel}
\usepackage{scalerel,amssymb}

\begin{document}
	\sloppy 
	\preprint{APS/123-QED}
	\allowdisplaybreaks
	\title{Predicting the Characteristics of Defect Transitions on Curved Surfaces}
	
	\author{Siddhansh Agarwal}
	\author{Sascha Hilgenfeldt}%
	\email{sascha@illinois.edu}
	\affiliation{Mechanical Science and Engineering, University of Illinois, Urbana-Champaign, Illinois 61801, USA}%


\begin{abstract}
	The energetically optimal position of lattice defects on intrinsically curved surfaces is a complex function of shape parameters. For open surfaces, a simple condition predicts the critical size for which a central disclination yields lower energy than a boundary disclination. In practice, this transition is modified by activation energies or more favorable intermediate defect positions. Here it is shown that these transition characteristics (continuous or discontinuous, first or second order) can also be inferred from analytical, general criteria evaluated from the surface shape. A universal scale of activation energy is found, and the criterion is generalized to predict  transition order as symmetries such as that of the shape are broken. The results give practical insight into structural transitions to disorder in many cellular materials of technological and biological importance.
\end{abstract}

 \sloppy
 \allowdisplaybreaks

 \maketitle

\section{Introduction}

A plethora of mechanical systems in nature and technological applications consist of interconnected units that form a thin shell or surface \cite{meng2014elastic,kohler2016stress,bausch2003grain,drenckhan2004demonstration,sknepnek2012buckling,nelson1995defects,roshal2020crystal}. The shape of these manifolds informs their function, and often intrinsic (Gaussian) curvature $K_G$ is required. Closed surfaces like viral capsids \cite{lidmar2003virus,vsiber2006buckling} or molecular cages \cite{bucher2018clathrin} have received much attention, but maybe even more common are open surfaces with a boundary, whether they are curved arrangements of microlenses \cite{chan2006fabricating}, the faceted eyes of insects \cite{kim2016hexagonal}, or topographically warped sheets of graphene or other metamaterials \cite{warner2012dislocation,paulose2015topological,silverberg2014using,peri2020experimental}. Capsids only become closed surfaces through assembly from open-surface states \cite{perlmutter2014pathways,perlmutter2015mechanisms}, and strategies to disrupt the assembly through, e.g. elastic frustration \cite{mendoza2020shape}, can be a powerful therapeutic tool \cite{klumpp2014capsid} which therefore need to be understood in detail. As is well known, non-zero $K_G$ on a lattice manifold is incompatible with a defect-free lattice; Euler's theorem applied to lattices translates into a statement about the topological charge $Q$, which on a triangulated lattice must be equal to 
\begin{align}
Q = \sum_{i=1}^V q_i = 6\chi,
\end{align}
where $q_i=6-c_i$ is the departure of the coordination number $c_i$ of the $i$th vertex from the ideal coordination number of a planar triangular lattice, and $\chi$ is the Euler characteristic. The most ubiquitous example of this is the presence of (a minimum of) 12 five-fold disclination defects on a soccer ball (with $\chi=2$). 
\begin{figure}
	\centering
    \includegraphics[width=0.5\textwidth]{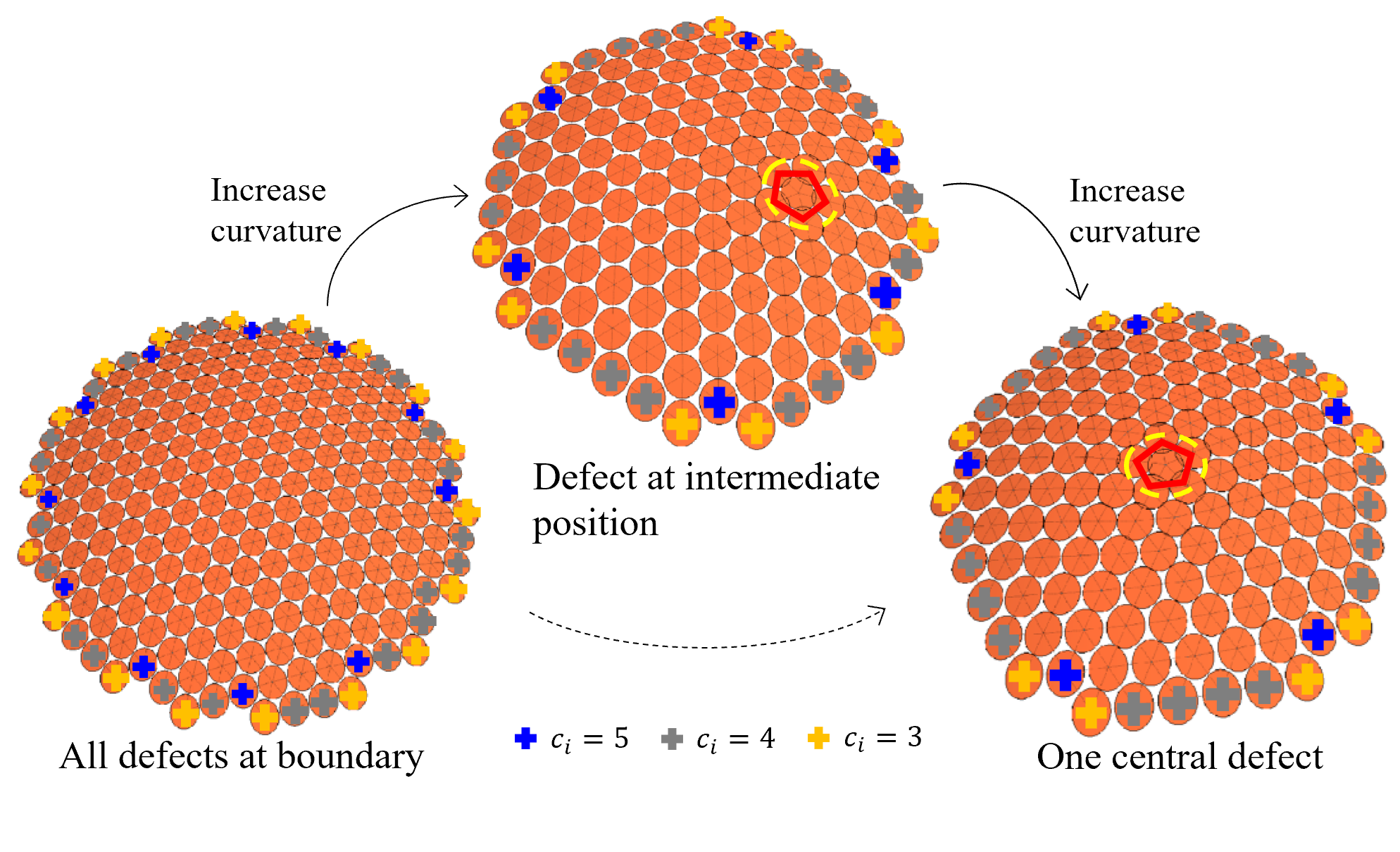}
	\caption{The elastic ground state of a weakly curved surface has all defects decorated at the boundary; upon increasing curvature, a disclination at the central apex eventually becomes favorable. This state is reached either continuously via intermediate defect positions (upper),  or discontinuously (lower). The path taken can be predicted by properties of surface shape.}\label{fig1_trans}
\end{figure}
Considerable work in the past years has focused not on the requirement of the total charge of defects, but on their positioning on the manifold \cite{bowick2000interacting,bowick2002crystalline,bowick2008bubble,giomi2007crystalline,vitelli2006crystallography,azadi2014emergent,azadi2016neutral,li2019}. This is particularly relevant for open surfaces, as here the magnitude of curvature controls the number of relevant defects visible on the bulk of the manifold: while Euler's theorem with $\chi=1$ still requires at least 6 disclinations, in a surface of small curvature they can be accommodated by the lattice at the boundary with minimal elastic energy penalty, so that the bulk of the surface remains defect-free. Only for sufficiently large curvature does it become energetically favorable for a single disclination to leave the boundary and migrate to the bulk (cf.\ Fig.~\ref{fig1_trans}). By extension, surfaces of varying Gaussian curvature have been shown to be preferentially populated by different numbers of defects \cite{irvine2010pleats,bowick2000interacting,bowick2004curvature}. 

For practical applications, it is highly desirable to have a unified description of this transition from a regular open surface to one with minimal disclination disorder. While local or integral Gaussian curvature can hint at preferred defect positions \cite{azadi2016neutral,irvine2010pleats}, very recently a general criterion was found that is applicable for a great variety of open-cap shape families: rather than a particular local value of $K_G$ or the value of the integrated Gaussian curvature over the entire surface, a particular {\em weighted integral} of $K_G$ needs to exceed a universal threshold in order for the defect migration to be energetically favorable \cite{agarwal2020}. For a certain surface shape, this translates to a critical size (cap extent) beyond which the elastic energy for the surface with a disclination at the apex of the cap is lower than that for the surface with all defects at the boundary (cf.\ Fig.~\ref{fig1_trans}).

However, what kind of transition is observed in practice upon increasing cap size depends not only on which configuration yields lower energy, but on the shape of the energy landscape: a transition may proceed continuously through intermediate defect positions representing energy minima, or discontinuously because an activation energy (energy barrier) needs to be overcome. These different {\em characteristics} of transition are illustrated in Fig.~\ref{fig1_trans}, and have previously been studied numerically for the special case of spherical shells  \cite{li2019}, where an analogy to first- and second-order phase transitions has been made. The possibility of symmetry-breaking ground states in continuous transitions is significant, as this alters the predictions of observable defect patterns. Conversely, a discontinuous transition ensures that symmetric defect positioning is the only realizable option. Symmetry is often a simplifying assumption for the building blocks in the simulation of virus capsid assembly \cite{grime2016coarse,hagan2014modeling}, and asymmetric placement of defects in an insect eye would disrupt optical regularity \cite{franceschini1972pupil}.

In the present work, we show that these transition characteristics can be reliably extracted for general shapes, via simple criteria involving the shape of the surfaces. Our present findings are applicable to a great variety of shapes and reveal universal properties of the energy landscape around transition.

In Sec.~\ref{sectheo} we sketch our rigorous theoretical framework for quantifying elastic energy on surfaces of revolution, and review results on the location of the transition in parameter space. Section~\ref{secchar} derives simple analytical criteria to determine the continuous or discontinuous character of transitions,  reveals universal properties in the energetic structure of states surrounding these transitions, and discusses how these energy landscapes shift when breaking the rotational symmetry of the manifold's mechanical properties. 
Section~\ref{secconcl} provides discussion and conclusions.

\section{Theoretical background}\label{sectheo}

In this study, we explore single disclination transition characteristics on a large set of bounded surfaces with rotational symmetry (unless stated otherwise) using linear elastic continuum theory. The surface geometry is imposed, i.e, we disregard  deformation degrees of freedom of the surface, whether elastic or through buckling, though these may play a role in other contexts \cite{lidmar2003virus,vsiber2006buckling,morozov2010assembly}. We focus on the transition from an energetically favorable defect-free surface (more precisely, all disclinations are located at the boundary) to one with a single disclination in the bulk. While the presence of dislocations can strongly affect such transitions \cite{azadi2014emergent,azadi2016neutral}, for large defect core energies dislocation distributions are prohibitively expensive energetically, while the single disclination provides a well-defined energy penalty scaling with the size of the surface \cite{seung1988defects}. Going beyond previous work \cite{agarwal2020}, we shall allow the disclination to occupy an arbitrary position on the surface in order to probe the energy landscape. 

\subsection{Full covariant formalism}\label{seccov}
We follow the covariant formalism of \citet{bowick2000interacting,giomi2007crystalline} with stress-free boundary conditions. The great accuracy of this approach for the present type of problem, and its relation to other formalisms detailed in \citet{li2019elasticity}, have been discussed in \citet{agarwal2020}. The elastic energy for an arbitrary disclination position ${\bf x}_D$ on a manifold $\mathbb{P}$ reads
\begin{align}
F_{\text{el}}(\mathbf{x}_D) = \frac{1}{2Y} \int \Gamma^2(\mathbf{x},\mathbf{x}_D) \mathrm{d} \mathbf{x},\label{Fel}
\end{align}
where the isotropic stress $\Gamma$ can be decomposed as
\begin{align}
\Gamma(\mathbf{x},\mathbf{x}_D) = -\Gamma_D(\mathbf{x},\mathbf{x}_D) -\Gamma_s(\mathbf{x}) + U(\mathbf{x},\mathbf{x}_D),\label{gamma2}
\end{align}
and
\begin{subequations}
\begin{align}
    \Gamma_D(\mathbf{x},\mathbf{x}_D) = - \frac{\pi}{3}Y G_L(\mathbf{x},\mathbf{x}_D),\label{gammaD}\\	\Gamma_S(\mathbf{x}) = Y\int K_G(\mathbf{y}) G_L(\mathbf{x},\mathbf{y})\mathrm{d} \mathbf{y}, \label{gammaS}
\end{align}
\end{subequations}
where
\begin{align}
    G_L(\mathbf{x};\mathbf{x}_D)= \frac{1}{2\pi}\log \bigg|\frac{z(\mathbf{x})-z(\mathbf{x}_D)}{1-z(\mathbf{x})\overline{z(\mathbf{x}_D)}}\bigg|\label{G_L}
\end{align}
is the Green's function of the covariant Laplace operator on $\mathbb{P}$ and $z(\mathbf{x})=\varrho(r)e^{i\phi}$ is a point on the unit disk on the complex plane while $\varrho(r)$ is the conformal radius of a map from the surface onto the unit disk. Here, \eqref{gammaD} is the contribution due to a disclination positioned arbitrarily at ${\bf x}_D$ while \eqref{gammaS} captures the screening effect of Gaussian curvature $K_G(\mathbf{x})$. The harmonic term $U({\bf x},{\bf x}_D)$ is determined by the boundary conditions at the rim of the surface; the sum of these three contributions is a fine balance of different effects. Balancing $\Gamma_D$ and $\Gamma_S$ represents a form of local curvature argument, where local Gaussian curvature compensates for a defect charge, while $U({\bf x},{\bf x}_D)$ introduces non-trivial boundary-dependent modifications. Going beyond \citet{giomi2007crystalline}, in the present work we perform the explicit computation of the harmonic function $U$ on the manifold $\mathbb{P}$ for arbitrary positioning of a disclination; one finds
\begin{align}
U({\bf x},{\bf x}_D) = -Y \int \mathrm{d} \mathbf{y} H(\mathbf{x},\mathbf{y})\left( \frac{\pi}{3} \delta(\mathbf{y},\mathbf{x}_D) -K_G(\mathbf{y})\right), \label{Ux}
\end{align}
where the harmonic kernel $H(\mathbf{x})$ is given by \cite{shimorin1997green}:
\begin{equation}
	H(\mathbf{x},\mathbf{y})=  -\frac{1}{2\pi} f_0(\varrho_y) - \sum_{n\geq 1} \frac{1}{\pi}\varrho_x^n \varrho_y^n \cos n(\phi_y-\phi_x) f_n(\varrho_y)
	\label{hkernel}
\end{equation}
This formalism relies on being able to explicitly find a conformal mapping from $\mathbb{P}$ onto the unit disk where points ${\bf x}, {\bf y}$ on the surface are mapped to complex numbers with moduli $\varrho_x$, $\varrho_y$ and arguments $\phi_x, \phi_y$, respectively (see Supplementary Information for details and the functional forms of $\varrho, f_n$).
The first term of (\ref{hkernel}) is a radially symmetric contribution due to the boundary conditions while the infinite sum acknowledges asymmetry introduced by, e.g., an intermediate position of the disclination. Accordingly, \eqref{Ux} can now be split into two terms: $ U(\mathbf{x},\mathbf{x}_D) = U_K({\bf x}) + U_{D}(\mathbf{x},\mathbf{x}_D)$, corresponding to the boundary contributions due to Gaussian curvature and the non-trivial disclination singularity, respectively, viz.
\begin{subequations}
	\begin{align}
	U_K(\mathbf{x})=& Y \int \mathrm{d} \mathbf{y} H(\mathbf{x},\mathbf{y})K_G(\mathbf{y})\label{Uk},\\
	U_{D}(\mathbf{x},\mathbf{x}_D)=&  \frac{f_0(\varrho_D)}{6}+\sum_{n\geq 1}\frac{(\varrho \varrho_D)^n}{3} f_n(\varrho_D) \cos n(\phi-\phi_D).\label{Udcl}
	\end{align}\label{U}\noindent
\end{subequations}
$U_K$ is a constant on rotationally symmetric surfaces, while $U_D(\mathbf{x},\mathbf{x}_D)$, which was not taken into account in earlier work \cite{giomi2007crystalline}, has to be computed with the series truncated at a large but finite $n$; in general both computations are executed numerically. Note that the infinite sum of (\ref{Udcl}) vanishes for a disclination at the boundary (where $f_{n \geq 1}=0$) as well as for a disclination at the center (where $\varrho_D=0$). For intermediate positions, as considered here, this term is generally non-zero. 

Taking (\ref{Fel}) together with (\ref{gamma2}) we have derived a rigorous covariant expression for the total energy  that can be evaluated for arbitrary positions of the disclination. This reads
\begin{align}
    F_{\text{el}}(\mathbf{x}_D) = \frac{1}{2Y}\int \left[(U_D-\Gamma_D) + (U_K-\Gamma_S)\right]^2 \mathrm{d}\mathbf{x}\,,\label{Felxd}
\end{align}
indicating that the energy penalty can be interpreted as a combination of mismatches between topological charges and harmonic contributions from disclinations and Gaussian curvature.

The difference of $F_{\text{el}}$ for a configuration with one defect at arbitrary $\mathbf{x}_D$ and $F_{\text{el}}$ for a configuration with only boundary defects (defect position $\mathbf{x}_D=\mathbf{x}_{b}$) is more explicitly
\begin{align}
    \Delta F_{\text{el}}(\mathbf{x}_D) = \frac{1}{2Y}\int (U_D-\Gamma_D)\left[(U_D-\Gamma_D) + 2(U_K-\Gamma_S)\right] \mathrm{d}\mathbf{x},
    \label{delfel}
\end{align}
because only $U_D$ and $\Gamma_D$ depend on $\mathbf{x}_D$.
Equation~(\ref{delfel}) is the generalization of the results of \citet{agarwal2020} to arbitrary disclination position $\mathbf{x}_D$. For the portion of the present work that considers radially symmetric surfaces, we can write $\Delta F_{\text{el}}(r_D)$. In the following, we scale out the (constant) material modulus, setting $Y=1$.

\subsection{Critical cap extent}\label{seccritcap}
The formalism above simplifies if the defect position is the apex of an axisymmetric cap surface ($\mathbf{x}_D=0$), which eliminates all angular dependences and makes $U_D$ and $U_K$ straight surface averages of $\Gamma_D$ and $\Gamma_S$, respectively. For a given surface shape, a critical cap radius $r_b=r_c$ can then be defined as the extent of the surface for which $\Delta F_{\text{el}}(0)=0$ according to (\ref{delfel}). 
\begin{figure}
	\centering
    \includegraphics[width=0.5\textwidth]{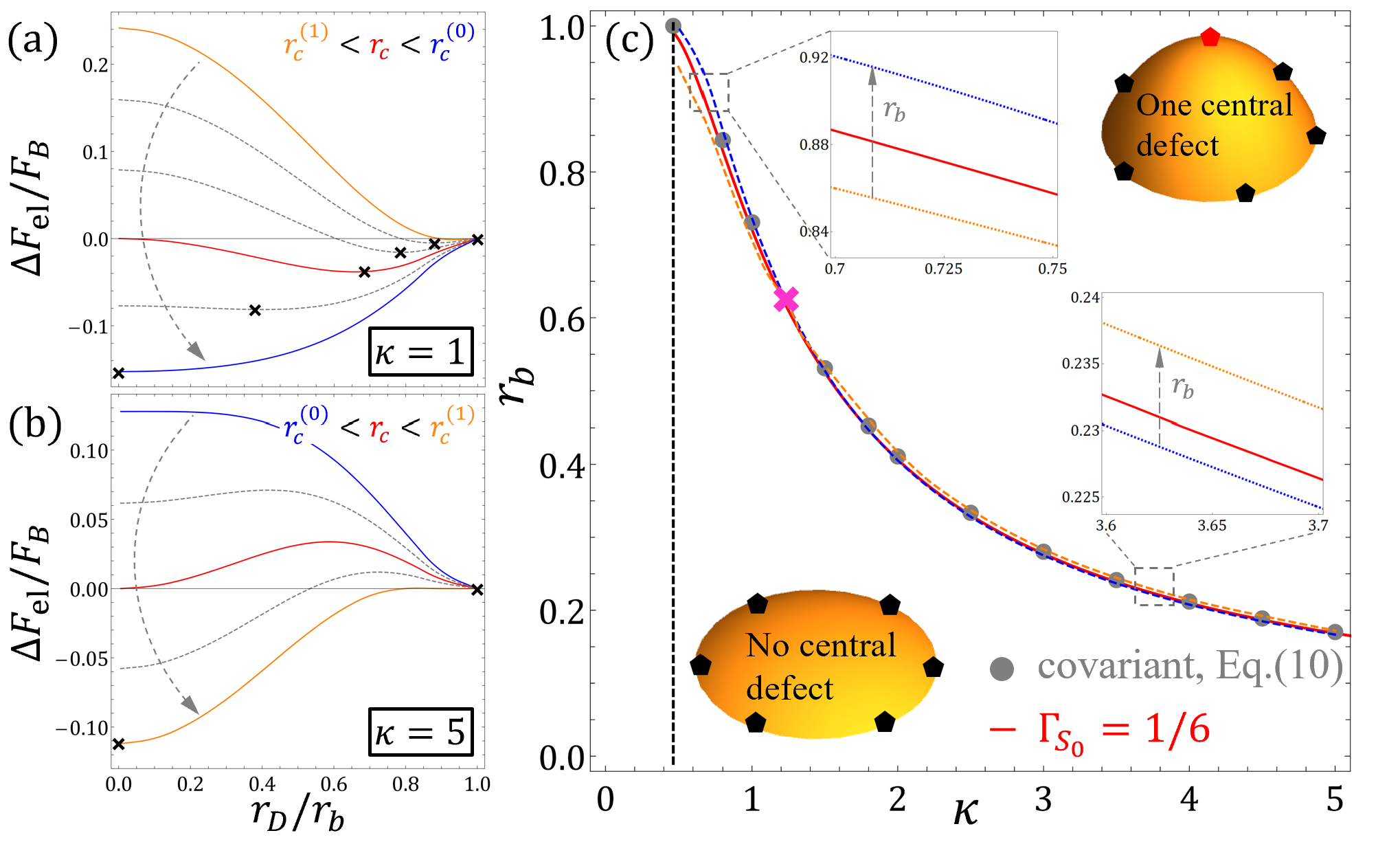}
	\caption{Normalized energy difference $\Delta F_{\text{el}}/F_B$ as a function of normalized defect position $r_D/r_b$ (varying cap extent $r_b$) for (a) Sphere ($\kappa=1$): the defect moves continuously from the boundary to the apex as $r_b$ is increased (the optimum intermediate positions are marked by crosses) -- it starts migrating at $r_b=r_c^{(1)}$ (orange curve) and reaches the center at $r_b=r_c^{(0)}$ (blue curve); (b) Prolate spheroid ($\kappa=5$): the defect migration is discontinuous and occurs abruptly once $r_b\geq r_c^{(1)}$ (orange curve).
	(c) Red curve is the boundary marking transition in the shape family of spheroids reproduced from \citet{agarwal2020}, while gray dots are numerically obtained roots using \eqref{delfel}. The orange-dashed $r_c^{(1)}$ and blue-dashed $r_c^{(0)}$ curves flanking the nominal red transition curve intersect it at a higher-order critical point $\kappa_h \approx 1.3$ (indicated by the magenta cross) and switch numerical order. Insets show a close-up of the curves for two distinct regimes of $\kappa$, on either side of the higher-order critical point. }\label{fig2_secondary}
\end{figure}
In \citet{agarwal2020} it was shown that, rather than numerically evaluating (\ref{delfel}), a simple criterion predicts $r_c$ with great accuracy, namely, 
\begin{align}
   \Gamma_S(0)\equiv \int\displaylimits_0^{r_{c}} K_G(r) \log \varrho(r) \sqrt{g}\,\mathrm{d}r = -\Gamma_{S_0}\, ,\label{gams0crit}
\end{align}
where $\Gamma_{S_0}\equiv 1/6$ is a universal constant. Thus, the Gaussian curvature weighted with the characteristic singularity $\log\varrho(r)$ of the defect stress governs the transition in parameter space. Eq.~(\ref{gams0crit}) yields $r_b=r_c$ as a function of shape parameters of the surface. In deriving this analytical criterion, an approximation was pursued that uses direct leading-order Taylor expansions of $\Gamma_D(r)$ and $g(r)$ around $r=0$, i.e., the small-slope approximations
\begin{align}
    \Gamma_{D}^{ss}(r,0) &= -\frac{1}{6} \log \left(r/r_{b}\right) ,\quad \sqrt{g}^{ss} =r,\label{gammadss}
\end{align}
but uses a non-local approximation for $\Gamma_S(r)$ that matches the values of this function at $r=0$ and the boundary position $r=r_b$, i.e.,
\begin{equation}
    \Gamma_{S}^{nl}(r) = \Gamma_S(0) \left(1-r^2/r_{b}^2\right),\label{gammasnl}
\end{equation}
where $\Gamma_S(0)$ is explicitly given by the integral in (\ref{gams0crit}), rather than using the small-slope expansion
\begin{equation}
    \Gamma_{S}^{ss}(r) = -\frac{1}{4} K_G(0) r_{b}^2 \left(1-r^2/r_{b}^2\right)\,.\label{gammasss}
\end{equation}
While the use of (\ref{gams0crit}) allows for accurate prediction of the transition in parameter space, superior to the small-slope approximation \cite{agarwal2020}, in this work we shall show that in order to predict the characteristics of the transition a further improvement in analytical theory is needed.

\section{Characteristics of the Transition}\label{secchar}

We parametrize general surfaces of revolution as $z=Z/L_r=f(r)$ with a radial length scale $L_r$, e.g.\ the equatorial radius of a spheroid. The dimensionless center curvature $f''(0)\equiv\kappa$ then represents a ratio of axial to radial length scales and is our primary parameter to vary shape within a shape family. For example, $f=\kappa \sqrt{1\pm r^2}$ describes spheroids ($-$) and hyperboloids ($+$) of various aspect ratio. We investigated many different shape families \cite{agarwal2020}, but confine ourselves to cap shapes with unique $z(r)$. For central defects ($r_D=0$) on spheroids, Fig.~\ref{fig2_secondary}c shows that the approximation (\ref{gams0crit}) (red line) predicts the transition to a central defect extremely accurately, compared with the numerical computation of $\Delta F_{\text{el}}(0)=0$ from (\ref{delfel}) (symbols). Caps of extent $r_b>r_c(\kappa)$ have lower energy with the disclination at the apex, while smaller caps have lower energy with all defects at the boundary. As pointed out above, this picture is complicated by the possibility of energy barriers or energy minima for intermediate disclination positions $0<r_D<r_b$.

\subsection{Covariant formalism -- effect of shape on secondary transition characteristics}\label{seccov_char}
The example of spheroids illustrates both of these characteristics of continuous and discontinuous transitions: Fig.~\ref{fig2_secondary}a plots the full covariant $\Delta F_{\text{el}}$ vs. defect position $r_D/r_b$ for spherical caps ($\kappa=1$), varying the cap extent $r_b$ through the critical value $r_c$. Here and in the following, we normalize energy differences by an elastic background energy scale 
\begin{align}
F_B\equiv \pi \kappa^4 r_{b}^6/384,
\end{align}
whose value is obtained by inserting the small-slope \eqref{gammasss} and $U_K =(1/A) \int \Gamma_S^{ss} dA = -(1/8)K_G(0)r_b^2$ into \eqref{Fel} for a defect free surface, i.e. $\Gamma_D = U_D=0$, and noting $K_G(0)=\kappa^2$.

The red line in Fig.~\ref{fig2_secondary}a (for $r_b=r_c$) shows that, while $\Delta F_{\text{el}}(0)=0$, the energy is even lower for an intermediate defect position. Increasing $r_b$ smoothly moves this optimum disclination position from the boundary to the apex (crosses). This process begins at $r_b=r_c^{(1)}<r_c$ determined by $\Delta F_{\text{el}}'(r_D=r_b)=0$ (orange curve) and ends at $r_b=r_c^{(0)}>r_c$ determined by $\Delta F_{\text{el}}'(r_D=0)=0$ (blue curve). Outside of this interval of $r_b$ values, $\Delta F_{\text{el}}(r_D)$ is monotonic. As noted in \citet{li2019}, spherical caps are thus an exponent of a continuous ("second-order") disorder transition.

By contrast, Fig.~\ref{fig2_secondary}b displays the energy landscapes for a prolate spheroid of $\kappa=5$. At $r_b=r_c$, the critical energy curve now shows a maximum, and this remains true for all $r_b$ in an interval $r_c^{(0)}<r_b<r_c^{(1)}$, where the $r_c^{(i)}$ are defined as above, but have switched numerical order. As a result, upon increasing $r_b$, an energy barrier prevents the disclination at the boundary from moving until $r_b>r_c^{(1)}>r_c$, when the defect abruptly jumps to the apex. This discontinuous transition thus requires a larger cap extent than $r_c$ or equivalently ``overcharging"  beyond the $q=1$ single disclination \cite{azadi2016neutral}.
\begin{figure}
	\centering
	 \includegraphics[width=0.5\textwidth]{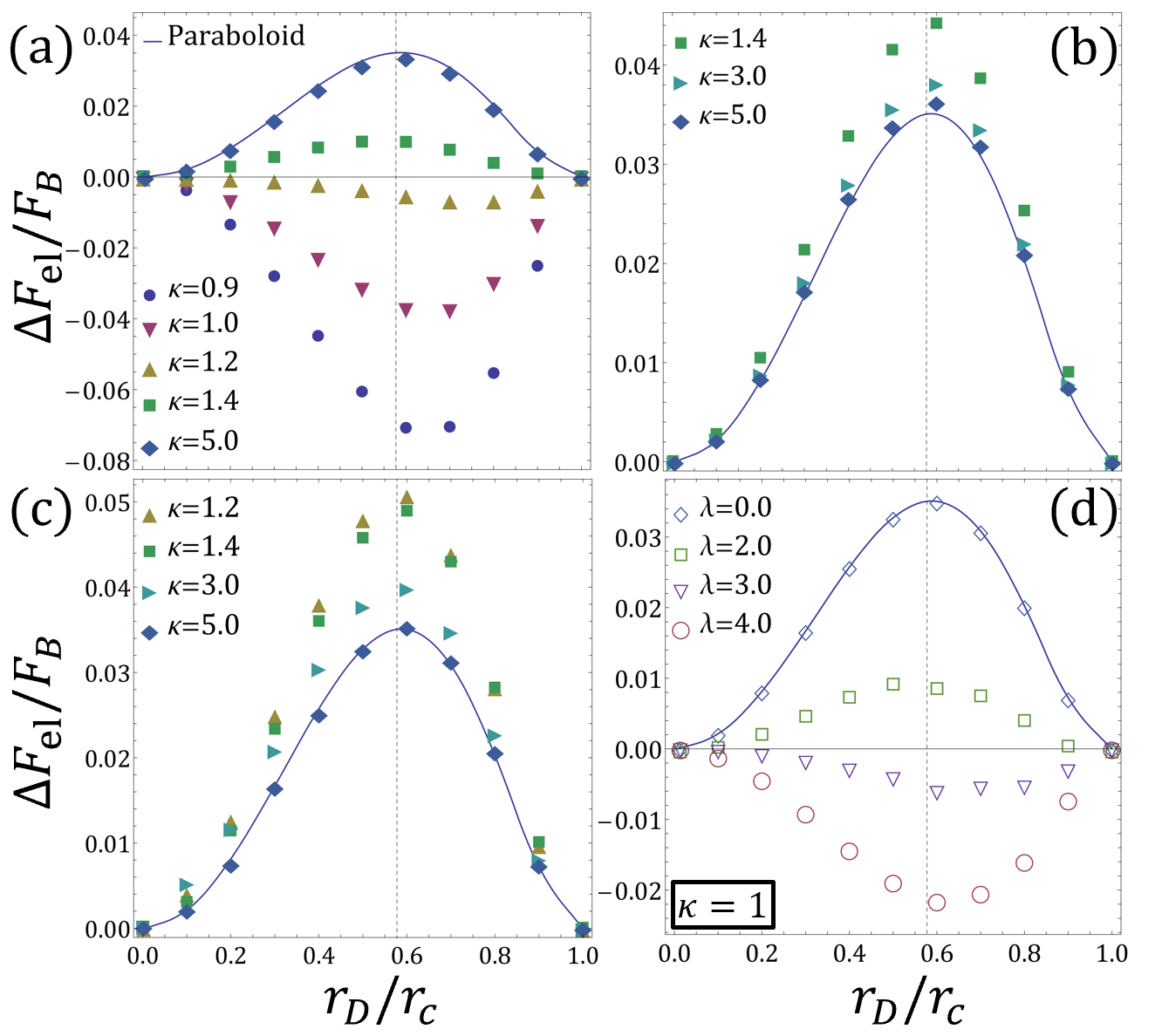}
	\caption{Normalized energy difference $\Delta F_{\text{el}}/F_B$ at transition ($r_b=r_c$) as a function of normalized defect position $r_D/r_c$ (varying $\kappa$) for (a) Spheroid: $f(r) = \kappa \sqrt{1-r^2}$, (b) Hyperboloid: $f(r) = \kappa \sqrt{1+r^2}$, (c) ``Bell-shaped" cap: $f(r) = \kappa/3 (1-r^2)^{3/2}$ and (varying $\lambda$) for (d) a prototypical higher-order surface: $f(r)=\kappa r^2/2+\lambda r^4/24$; the character of the transition is continuous for $\lambda/\kappa \gtrsim 3$. The large $\kappa$ asymptote for all shapes has a common energy barrier -- identical to that of a Paraboloid (indicated by solid curves). The location of the intermediate extremum is obtained by solving \eqref{delfelprime} and is approximately $r_{D,m}=r_c/\sqrt{3}$ (indicated by dashed vertical lines).}\label{fig3_crit}
\end{figure}
In Fig.~\ref{fig2_secondary}c, we plot the $\kappa$-dependent values of $r_c^{(0)}$ (blue-dashed) and $r_c^{(1)}$ (orange-dashed) for the entire family of spheroidal caps. They flank the red line of the nominal transition criterion $\Delta F_{\text{el}}(0)=0$ closely and intersect it at a higher-order critical point where $r_c^{(0)}$  and $r_c^{(1)}$ switch order and, therefore, continuous transitions become discontinuous.
The range of cap sizes over which the transition character manifests itself (for a given $\kappa$) is relatively small, as illustrated in the two insets of Fig.~\ref{fig2_secondary}(c).

Can we observe such features more generally, going beyond spheroids? In Fig.~\ref{fig3_crit},
we take a closer look at the shape and scale of the energy landscape at critical extent $r_b=r_c$
for four different shape families, varying $\kappa=f''(0)$ for the first three and the fourth apical derivative $\lambda = f^{(4)}(0)$ (keeping $\kappa$ fixed) for the last one. 
Figure~\ref{fig3_crit}a again illustrates the transition from continuous to discontinuous in spheroids, and pinpoints the higher-order critical point at $\kappa\approx 1.3$. Hyperboloids (a shape family with potentially infinite cap extent, Fig.~\ref{fig3_crit}b) and a family of "bell-shaped" surfaces whose Gaussian curvature changes sign on the surface (Fig.~\ref{fig3_crit}c), by contrast, never show continuous transitions. For large $\kappa$, however, the normalized energy difference $\Delta F_{\text{el}}/F_B$  for all shapes approaches a common energy barrier. Figure~\ref{fig3_crit}d shows that the transition character can be changed at constant $\kappa$ by changing the value of the fourth apical derivative $\lambda$.

This is an indication that higher apical derivatives of the surface shape play a central role here. Indeed, when solving this problem for the unique surface without higher derivatives (the paraboloid), the result accurately depicts the common asymptotic shape of the energy landscape (solid lines in Fig.~\ref{fig3_crit}).

The numerical integration of (\ref{delfel}) yields these results, but gives little physical insight. When do we expect continuous vs.\ discontinuous transitions? Is the scale of the energy landscape (a few percent of the normalization value $F_B$) indeed universal, as suggested by these examples? Exactly which features of the surface are determinants of the transition character? To answer these questions, we turn to analytical approximations.

\subsection{Analytical theory: non-local approximation}\label{secanalytica}
\begin{figure}
    \centering
	 \includegraphics[width=0.5\textwidth]{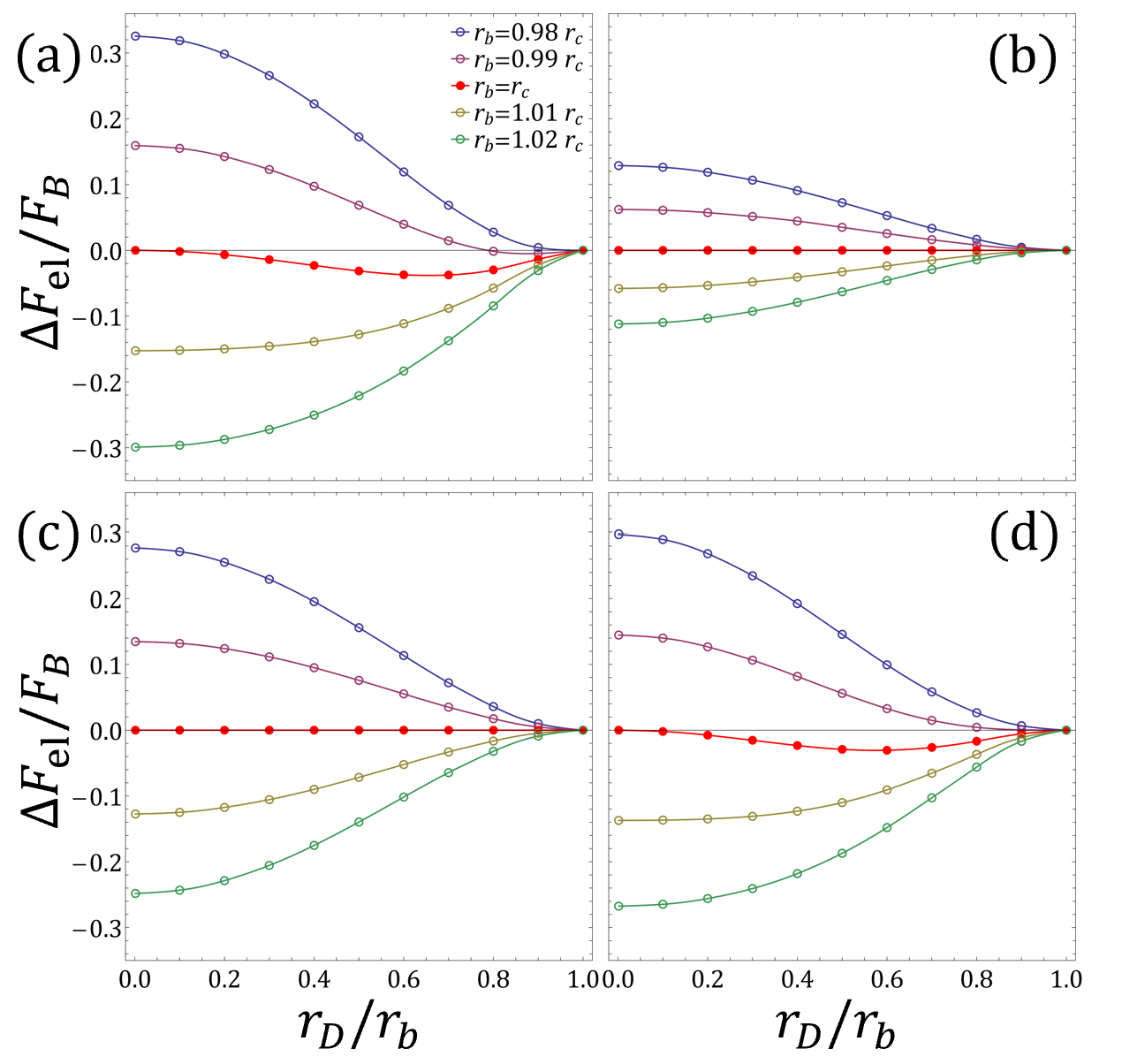}
	\caption{Normalized energy difference $\Delta F_{\text{el}}/F_B$ for a sphere ($\kappa=1$), comparing different approaches (varying $r_b$ around $r_c$): (a) full covariant Eq. \eqref{delfel} (numerical), (b) small-slope (cf. \citet{azadi2016neutral}), (c) non-local formalism from \citet{agarwal2020}, (d) non-local formalism of Eq. \eqref{delfel_nl}. Only the latter, non-local analytical approach captures the characteristics of the transition. }\label{fig4_compare_theory}
\end{figure}

As detailed in section~\ref{seccritcap}, the accurate prediction of the transition line in Fig.~\ref{fig2_secondary}c was not possible with a small-slope Taylor expansion, but needed the non-local improvement (\ref{gammasnl}) \cite{agarwal2020}. Using either $\Gamma_S^{ss}$ or $\Gamma_S^{nl}$, and the resulting changes in $U_K$, while maintaining the approximations (\ref{gammadss}), $\Delta F_{\text{el}}(r_D)$ becomes analytically tractable. Figure~\ref{fig4_compare_theory}a reproduces the continuous structure of the transition for $r_b$ around $r_c$ for spherical caps according to the rigorous numerical computation. Figures~\ref{fig4_compare_theory}b and c show the analogous results for the small-slope and non-local approximations, respectively. Intriguingly, neither of these approaches produces any secondary characteristics at all -- the energy difference at transition is flat, and monotonic at every other value of $r_b$. Thus, even though the non-local approximation reproduces the spread of energy values more accurately than the small-slope model, it does not probe the shape properties of the surface that are responsible for the order of the transition.

We now describe a minimal model that captures the dominant effects of the secondary transition character. Noting that a more accurate representation of the intrinsic isotropic stress $\Gamma_S(r)$ was crucial for predicting the transition location, we improve further on the approximation (\ref{gammasnl}): in addition to matching the function value and first derivative at $r=0$ and function value at $r=r_b$, we now require a match of the second derivative at the apex. By symmetry, this requires a fourth-order polynomial in $r$, namely,
\begin{align}
     \Gamma_{S}^{nl}(r) = &-\frac{\Gamma_S''(0)r_b^2}{2} \left(1-\frac{r^2}{r_{b}^2}\right)\nonumber\\
     &+\left(\Gamma_S(0)+\frac{\Gamma_S''(0)r_b^2}{2}\right)\left(1-\frac{r^4}{r_{b}^4}\right),
\end{align} \label{gammasnlr4}
with the full rigorous expressions
\begin{align}
   \Gamma_S(0)\equiv& \int\displaylimits_0^{r_{b}} K_G(r) \log\varrho(r) \sqrt{g}\,\mathrm{d}r,\, \Gamma_S''(0)=\frac{K_G(0)}{2}=\frac{\kappa^2}{2}\,. \label{gams0}
\end{align}
We use \eqref{gammaD} and \eqref{G_L} for arbitrary defect position $\mathbf{x}_D$ but employ the small-slope $\varrho^{ss}(r)=r/r_b$ and $\sqrt{g}^{ss}=r$. Inserting these into \eqref{delfel} and \eqref{U}, the energy integral can be executed analytically (see Supplementary Information for details) and results in the following closed form expression:
\begin{align}
    \Delta F_{\text{el}}(r_D) = \frac{\pi r_b^2}{864} \left(1-\frac{r_D^2}{r_b^2}\right)^2 \bigg[&3 + 8\left(2+ \frac{r_D^2}{r_b^2}\right)\Gamma_S(0)\nonumber\\
    &- \left(1-4\frac{r_D^2}{r_b^2}\right)\frac{\kappa^2 r_b^2}{2} \bigg].\label{delfel_nl}
\end{align}
The transition threshold is defined as the extent $r_b$ at which $\Delta F_{\text{el}}(0)=0$ and \eqref{delfel_nl} results in 
\begin{align}
      \Gamma_S(0) = -3/16 +\kappa^2 r_c^2/32 \label{trans_crit}
\end{align}
as the transition criterion within this level of approximation. The extremum of the energy landscape is obtained from
\begin{align}
    \Delta F_{\text{el}}'(r_D) = \frac{\pi r_D}{72} \left(\frac{r_D^2}{r_b^2}-1\right) &\bigg[1 + 4 \left(1+\frac{r_D^2}{r_b^2}\right)\Gamma_S(0)\nonumber\\
    &- \left( 1-2\frac{r_D^2}{r_b^2}\right)\frac{\kappa^2}{2}r_b^2\bigg] \label{delfelprime}
\end{align}
by setting $\Delta F_{\text{el}}'(r_D)\vert_{r_b=r_c} =0$. Inserting the transition criterion \eqref{trans_crit} leads to a straightforward non-trivial solution for the position of the minimum or maximum of the energy landscape $r_{D,m} = r_c/\sqrt{3}$. This result is indicated by dashed vertical lines in Fig.~\ref{fig3_crit} and in good agreement for the vast majority of shapes. Inserting $r_{D,m}$ together with \eqref{trans_crit} into \eqref{delfel_nl} yields the scale of the secondary structure. Normalizing by $F_B$, one obtains
\begin{align}
    \Delta F_{\text{el}}^{\text{sec}}/F_B = \frac{4}{81\kappa^4 r_c^4}\left(-2 + 3\kappa^2 r_c^2 \right)\label{delfelsec}
\end{align}
This rigorously shows why a small-slope formalism is incapable of predicting secondary transition structure (since $\kappa^2r_c^2=2/3$ in this approximation \cite{azadi2016neutral}). Further analytical progress can be made by determining $r_c(\kappa)$ at transition within the present approximation. 
By symmetry, the large $\kappa$ expansion of $r_c$ for rotationally symmetric surfaces is $r_c = a/\kappa + b/\kappa^3 + \dots$ (cf.\ \citet{agarwal2020}). Inserting this into \eqref{trans_crit} and Taylor-expanding both sides for $\kappa \to \infty$ allows us to solve for the coefficients $a$, $b$, $\dots,$ order-wise. With the definition of $\Gamma_{S}(0)$, this reads 
\begin{align}
    \Gamma_{S}(0)   = &1-\sqrt{1+a^2}+\log \left[\frac{\left(1+\sqrt{1+a^2}\right)}{2} \right]+ \frac{b(1-\sqrt{a^2+1})}{a \kappa^2}\nonumber\\
    &+\frac{\left(-a^4+a^2-2 \sqrt{a^2+1}+2\right) (\lambda/\kappa)}{18 \sqrt{a^2+1} \kappa^2} +\dots \label{gammas0exp}
\end{align}
Finally, expanding the RHS of \eqref{trans_crit} and equating with \eqref{gammas0exp} we obtain
$a\approx 0.844, \, b\approx -0.042\lambda/\kappa$. Note that a generic surface (with one dominant radial length scale) will have $\lambda = \mathcal{O}(\kappa)$ and $b=\mathcal{O}(1)$. Inserting into a large $\kappa$ expansion of \eqref{delfelsec}, we obtain:
\begin{align}
    \Delta F_{\text{el}}^{sec}/F_B &=  \frac{4(-2+3a^2)}{81a^4} + \frac{8(4-3a^2)b}{81a^5\kappa^2}+\dots\nonumber\\
    &\approx 0.0135 - 0.018 \frac{\lambda}{\kappa^3} +\dots
\end{align}
This suggests a universal scale for the energy barrier at transition for surfaces with large central curvature, consistent with the observations of  Fig.~\ref{fig3_crit}. For $\kappa\to\infty$, higher derivatives are negligible and the behavior mimics that of a paraboloid. Furthermore, the result predicts a change from an energy maximum to a minimum at transition (a higher-order critical point), only if the quantity $\lambda/\kappa$ is positive.
Note that for spheroids $\lambda/\kappa=3 >0$, whereas for the hyperboloids or ``bell" shapes of Fig.~\ref{fig3_crit}bc $\lambda/\kappa=-3<0$. Likewise, increasing $\lambda$ from zero should lead to the development of an energy minimum and thus a continuous transition, as shown in Fig.~\ref{fig3_crit}d.

The formalism outlined here also settles the question of the range of the secondary structure effects described in Sec.~\ref{seccov_char} i.e., the values of $r_c^{(0)}$ and $r_c^{(1)}$. Using the defining criteria $\Delta F'_{\text{el}}(r_D=0)\vert_{r_b=r_c^{(0)}}=0$ and $\Delta F'_{\text{el}}(r_D=r_b)\vert_{r_b=r_c^{(1)}}=0$ in \eqref{delfelprime} results in the following implicit equations:
\begin{align}
    \Gamma_S(0) = -\frac{1}{8} +\frac{(\kappa r_c^{(0)})^2}{16}, \quad \Gamma_S(0) = -\frac{1}{4} +\frac{(\kappa r_c^{(1)})^2}{8} \label{sec_crit}\,,
\end{align}
respectively. As before, we insert the large $\kappa$ expansion $r_c^{(i)} = a/\kappa + b/\kappa^3 + \dots$, Taylor expand and determine the respective coefficients $a,b,\dots$ to finally obtain
\begin{align}
    (r_c^{(1)}-r_c^{(0)})/r_c \approx 0.038 - 0.07\lambda/\kappa^3
\end{align}
This suggests that the range of the secondary features is about $4\%$ relative to the critical $r_c$ for surfaces with large central curvature -- again, the large $\kappa$ limit universally asymptotes to that of a paraboloid. On the other hand, shapes with small $\kappa$ could have larger ranges, but are not accurately described by this approach. Empirically, we see for all shape families analyzed that the range of secondary characteristics remains of this order up to the smallest $\kappa$ consistent with unique $f(r)$ shapes.

All of these predictions are in complete qualitative agreement with the numerical computations, while quantitative discrepancies in $\Delta F_{\text{el}}^{\text{sec}}$ can be systematically improved upon employing higher order approximations of the type that also improve the predictive error of $r_c(\kappa)$ \cite{agarwal2020}. Matching the third derivative of the metric $\sqrt{g}$ at $r=0$ yields
\begin{align}
    \sqrt{g} = r + \frac{\kappa^2}{2}r^3.
    \label{sqrtg3}
\end{align}
With only this modification ($\Gamma_D$ remains unchanged from \eqref{gammadss} to permit analytical evaluation of the integrals) we go through the same steps outlined above and, while the explicit expressions are more complicated, we obtain a slightly different transition criterion $r_c \approx 0.829/\kappa -0.039(\lambda/\kappa)/\kappa^3$ and an altered secondary energy,
\begin{align}
    \Delta F_{\text{el}}^{\text{sec}}/F_B \approx 0.052 - 0.017 \frac{\lambda}{\kappa^3} +\dots,
    \label{felsecimproved}
\end{align}
which more closely approximates the empirical universal barrier height for $\kappa\to\infty$, which is $\Delta F_{\text{el}}^{\text{sec}}/F_B \approx 0.034$ in the full covariant computation. Eq.~(\ref{felsecimproved}) predicts that the higher-order critical point is given by $\lambda_h \approx 3\kappa_h^3$. For spheroids, this locates the transition from continuous to discontinuous behavior at $\kappa_h\approx 1$, again in good agreement with the covariant computation ($\kappa_h\approx 1.3$). The remaining quantitative discrepancies can be systematically alleviated by including higher order terms in the non-local expansions, although not all integrals may be tractable analytically. 

\subsection{Non-local vs. local approximations}\label{secnl}

The results from the previous subsection would suggest that only local information about the surface (such as $\kappa$ or $\lambda$) is sufficient to predict the nature of the transition, even though we employed a non-local formalism with the full $\Gamma_S(0)$. This raises the question whether a local higher-order small-slope formalism would yield similar quantitative agreement with the full covariant formalism. In order to test this, we instead employ a local fourth-order expansion of $\Gamma_S(r)$ that reads:
\begin{align}
     \Gamma_{S}^{l}(r) = -\frac{\kappa^2 r_b^2}{4} \left(1-\frac{r^2}{r_{b}^2}\right)+\frac{\kappa^4 r_b^4}{96}\left(3-4\frac{\lambda}{\kappa^3}\right)\left(1-\frac{r^4}{r_{b}^4}\right),\label{gammaslr4}
\end{align} 
where we now use the small-slope expansion for $\Gamma_{S}(0)=-\kappa^2r_b^2/4+(3-4\lambda/\kappa)\kappa^4r_b^4/96 +\dots$, retaining terms up to $\mathcal{O}(r^4)$. Together with \eqref{sqrtg3} ($\Gamma_D$ remains unmodified from \eqref{gammadss}), we go through the same steps as outlined in the previous subsection, i.e. we evaluate the energy integral analytically, determine the transition criterion at this level of approximation, compute the position of the extremum and finally insert all these into $\Delta F_{\text{el}}/F_B$ to obtain the scale of the normalized secondary structure, which reads
\begin{align}
    \Delta F_{\text{el}}^{\text{sec}}/F_B \approx 0.057 - 0.021 \frac{\lambda}{\kappa^3} +\dots.\label{felseclocal}
\end{align}
\begin{figure}
	\centering
	 \includegraphics[width=0.5\textwidth]{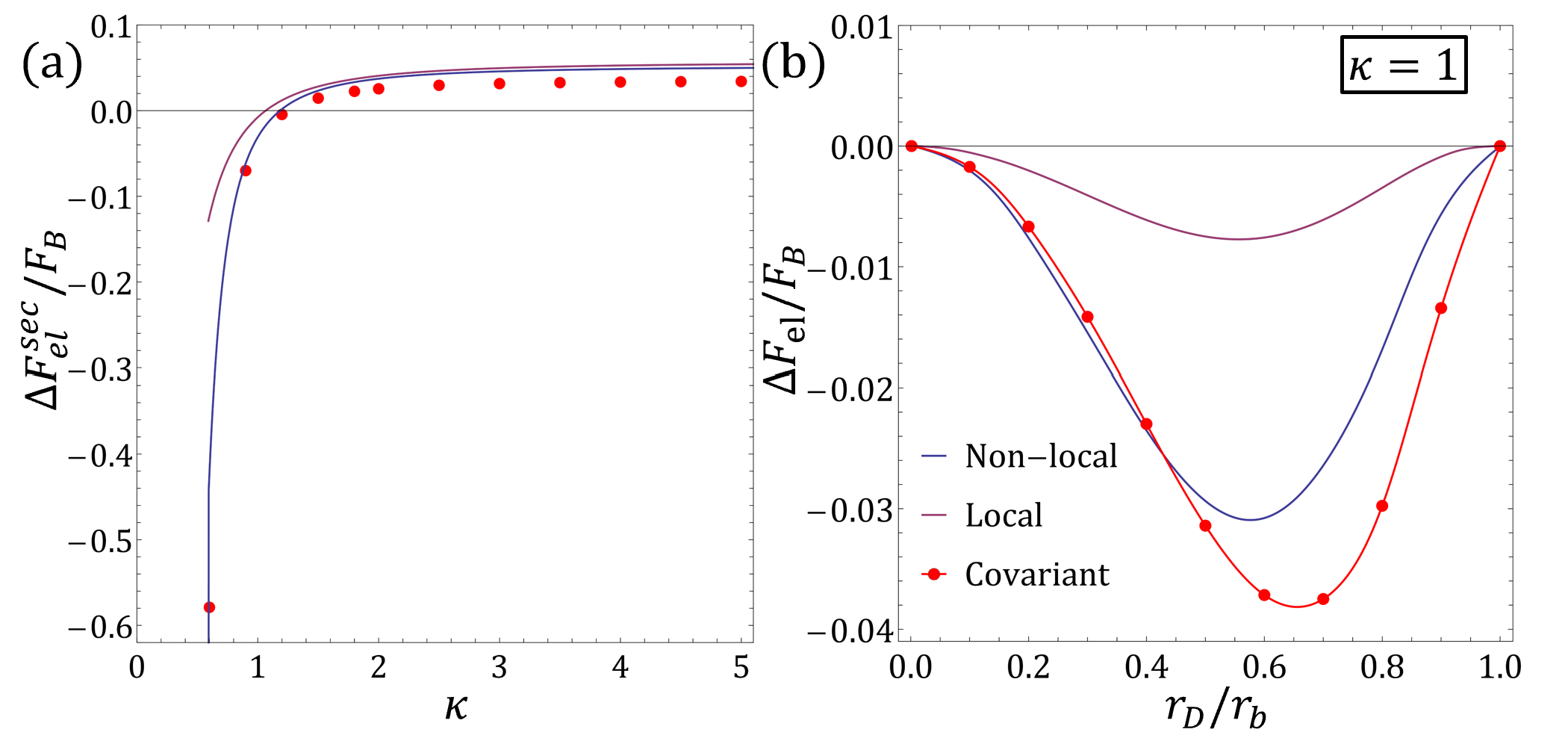}
	\caption{(a) Scale of the normalized secondary energy structure $\Delta F_{\text{el}}^{\text{sec}}/F_B$ at transition: the red dots were obtained numerically by integrating the full covariant equation. While both non-local approximation (blue) and a higher order $\mathcal{O}(r^4)$ local approximation (magenta) have the same large $\kappa$ asymptote, the local approach has large unphysical deviations from the covariant formalism for $\kappa \lesssim 1$. (b) Plotting the normalized energy difference $\Delta F_{\text{el}}/F_B$ at transition for a sphere ($\kappa=1$) showcases this large quantitative discrepancy in the secondary energy structure.}\label{fig5_nl}
\end{figure}
We plot in Fig.~\ref{fig5_nl}(a) the full covariant $\Delta F_{\text{el}}^{\text{sec}}/F_B$ obtained by numerically integrating \eqref{delfel} (indicated by red dots) and compare the local approximation with the non-local one for the shape family of spheroids as our prototypical example. The large central curvature ($\kappa\to \infty$) limit for both converges to approximately the same asymptote with similar deviations from the covariant prediction (cf. \eqref{felseclocal} and \eqref{felsecimproved}), however differences between the two become significant for surfaces with smaller values of $\kappa$ that are arguably of more practical relevance \cite{agarwal2020}. Fig.~\ref{fig5_nl}(b) exemplifies the significant discrepancy for the particular case of $\kappa=1$ (sphere) -- while the local approximation qualitatively captures the behavior at transition it greatly underestimates the magnitude of the energy minimum and the disagreement only gets worse for surfaces with $\kappa<1$. Therefore, analogous to the transition criterion from our previous work \cite{agarwal2020}, quantitatively describing the scale of the secondary structure around transition requires non-local information from the entire surface for shapes with $\kappa\sim \mathcal{O}(1)$.

We now turn our attention to the question of how robust this secondary-structure behavior is. In particular, the small range of cap sizes over which it occurs suggests it may be qualitatively altered by even slight deviations from the radial symmetry of the surface shape.

\subsection{Rotational symmetry breaking}\label{secshape}

In general, the boundary of open surfaces will not obey radial symmetry; the {\em Drosophila} eye, for example, is well approximated by an ellipsoidal cap with an elliptical boundary \cite{hilgenfeldt_unpublished}. While a fully covariant formalism in such a general scenario might be feasible numerically, we are not aware of any previous work that studies this role of rotational symmetry breaking. In order to estimate the magnitude of the effect of breaking boundary shape symmetry on the energy landscape, we shall utilize the small-slope approximation; as we have shown above, this does not, by itself, lead to any secondary structure of the transition (cf.\ Fig.~\ref{fig4_compare_theory}b). As we have furthermore shown that the energy scale of intrinsic secondary structures is bounded at least for large central curvature $\kappa$, we can give a quantitative estimate of whether symmetry breaking will change this structure.

We apply a leading-order shape perturbation to a general surface of revolution, generating an ellipse from the circular boundary by stretching/contracting perpendicular axes by an amount $\epsilon$. Thus, 
\begin{align}
x=(1+\epsilon)r\cos \phi,\quad y=(1-\epsilon)r \sin \phi, \quad z= f(r)
\end{align}
with $\epsilon\ll 1$, while the metric tensor in the small-slope limit is 
\begin{align}
g_{ij}=\begin{bmatrix} 
	1 + \epsilon(2+\epsilon)\cos 2\phi & -2r\epsilon \sin 2\phi \\
	-2r\epsilon \sin 2\phi & r^2\left(1 + \epsilon(\epsilon-2)\cos 2\phi\right)
\end{bmatrix}.\label{gijss}
\end{align}
Here, $\sqrt{g}=r(1-\epsilon^2)$ and eccentricity $e=\sqrt{1-\frac{(1-\epsilon)^2}{(1+\epsilon)^2}}=2\sqrt{\epsilon} + \mathcal{O}(\epsilon^{3/2})$, while the small-slope Gaussian curvature is constant to $\mathcal{O}(\epsilon)$ and is given by 
\begin{align}
K_G(r,\phi) = \kappa^2 +\mathcal{O}\left(\epsilon^2\right)\,.
\end{align}
The elliptical boundary of a section cut parallel to the $xy$-plane is given by
\begin{align}
r(\phi) = \frac{r_b(1-\epsilon^2)}{\sqrt{1+\epsilon(\epsilon-2)\cos 2\phi}} \approx r_b(1+\epsilon \cos 2\phi).
\label{ellipse}
\end{align}
Breaking rotational symmetry of the surface leads to coupling of stresses in the radial and azimuthal directions. Therefore, the isotropic stresses are not as easily computed as in the previous subsections. Instead, we start by observing that the Airy stress function $\chi$ satisfies the following equation in polar coordinates \cite{azadi2016neutral}:
\begin{align}
\nabla^4_\perp \chi &=  \frac{\pi}{3}\frac{\delta(r-r_D)\delta(\phi-\phi_D)}{\sqrt{g}}-K_G(r,\phi) \label{biharmperp}
\end{align}
subject to a zero normal stress boundary condition and regularity conditions at $r=0$. Here, $\nabla^2_\perp$ is the 2D Laplace-Beltrami operator in the small-slope limit and is given by $\nabla^2_\perp f= \frac{1}{\sqrt{g}}\partial_i\left(\sqrt{|g|}g^{ij}\partial_j f\right)$ using the small-slope limit \eqref{gijss}. Here, $g^{ij}$ is the inverse of the metric tensor such that $g^{ij}g_{jk}=\delta^i_k$.
The Airy function is related to the stress components in the usual way,
\begin{equation}
   \sigma_{rr} = \frac{1}{r}\frac{\partial \chi}{\partial r} + \frac{1}{r^2} \frac{\partial^2 \chi}{\partial \phi^2},\, \sigma_{r\phi}= -\frac{\partial}{\partial r}\left(\frac{1}{r} \frac{\partial \chi}{\partial \phi}\right),\, \sigma_{\phi\phi} = \frac{\partial^2 \chi}{\partial r^2}.
\end{equation}
Analogous to the case of a circular boundary, we impose vanishing 
normal stress, i.e. $\bm{\sigma} \cdot \hat{\bm{n}}=0$ on (\ref{biharmperp}), where $\hat{\bm{n}} = n_r \hat{\bm{e}}_r+ n_\phi \hat{\bm{e}}_\phi$ is the normal vector to the elliptical boundary. Therefore, one obtains two scalar equations:
\begin{align}
\bigg[\left(n_r \sigma_{rr} + n_\phi \sigma_{r\phi}\right)\hat{\bm{e}}_r + \left(n_r \sigma_{r\phi} + n_\phi \sigma_{\phi\phi}\right)\hat{\bm{e}}_\phi\bigg]_{r=r_b(\phi)} =0\,.
\end{align}
The total in-plane elastic energy for a surface with stress-free boundary and metric $g$ in terms of the stress components (assuming a linear constitutive relation) is given by \cite{giomi2007crystalline,azadi2016neutral}:
\begin{align}
F_{\text{el}} = \frac{1}{2}\int  \Gamma(r,\phi)^2 dA=\frac{1}{2}\int  \left(\sigma_{rr}+\sigma_{\phi\phi}\right)^2 dA,\label{Fel_ellipse}
\end{align}
where $\Gamma = \sigma_{rr} +\sigma_{\phi\phi}$ is the trace of the stress tensor.

Writing $\chi$, and thus $\sigma_{ij}$, as a Fourier series expansion in $(\cos n\phi, \sin n\phi)$ and consistently expanding the stress components as well as $\nabla^2_\perp$ and the boundary condition in powers of $\epsilon$, we evaluate the elastic energy up to $\mathcal{O}(\epsilon)$ (see Supplemental Information for details). Inserting the expansion $\Gamma =  \Gamma_0+\epsilon \Gamma_1$ into \eqref{Fel_ellipse}, the elastic energy can be cast explicitly as
\begin{align}
F_{\text{el}} =\frac{1}{2}\int\limits_0^{2\pi}\int\limits_0^{r_b}\left( \Gamma_0^2 + 2\epsilon\Gamma_0 \Gamma_1 \right)r drd\phi\, +\mathcal{O}(\epsilon^2).
\end{align}
In terms of Fourier components, it is evident that only the squares of the modes will contribute to the energy while the cross terms will integrate out to zero.

Systematically evaluating the $\mathcal{O}(1)$ and $\mathcal{O}(\epsilon)$ contributions to the stress tensor, we obtain an angular correction to $\Delta F_{\text{el}}$. The final expression reads
\begin{align}
    \frac{\Delta F_{\text{el}}(\overline{r}_D)}{F_B}&\approx \frac{2\left(2-3\kappa^2r_b^2\right)}{3\kappa^4 r_b^4}\left(1-\overline{r}_D^2\right)^2 \nonumber\\
    &+\frac{4}{3}\epsilon\overline{r}_D^2\cos 2\phi_D \left(\frac{-1+2\overline{r}_D^2-3\overline{r}_D^4+2\overline{r}_D^6-4\log\overline{r}_D}{\kappa^4 r_b^4}\right),\label{DelFel_ellipse}
\end{align}
where $\overline{r}_D=r_D/r_b$. The leading order is the small-slope energy expression for a symmetric cap obtained, e.g., by \citet{azadi2016neutral}, while the next term represents the leading effect of the shape perturbation and depends, in addition to $r_D$, on the angular position $\phi_D$ of the defect. 

\begin{figure}
	\centering
	\includegraphics[width=0.5\textwidth]{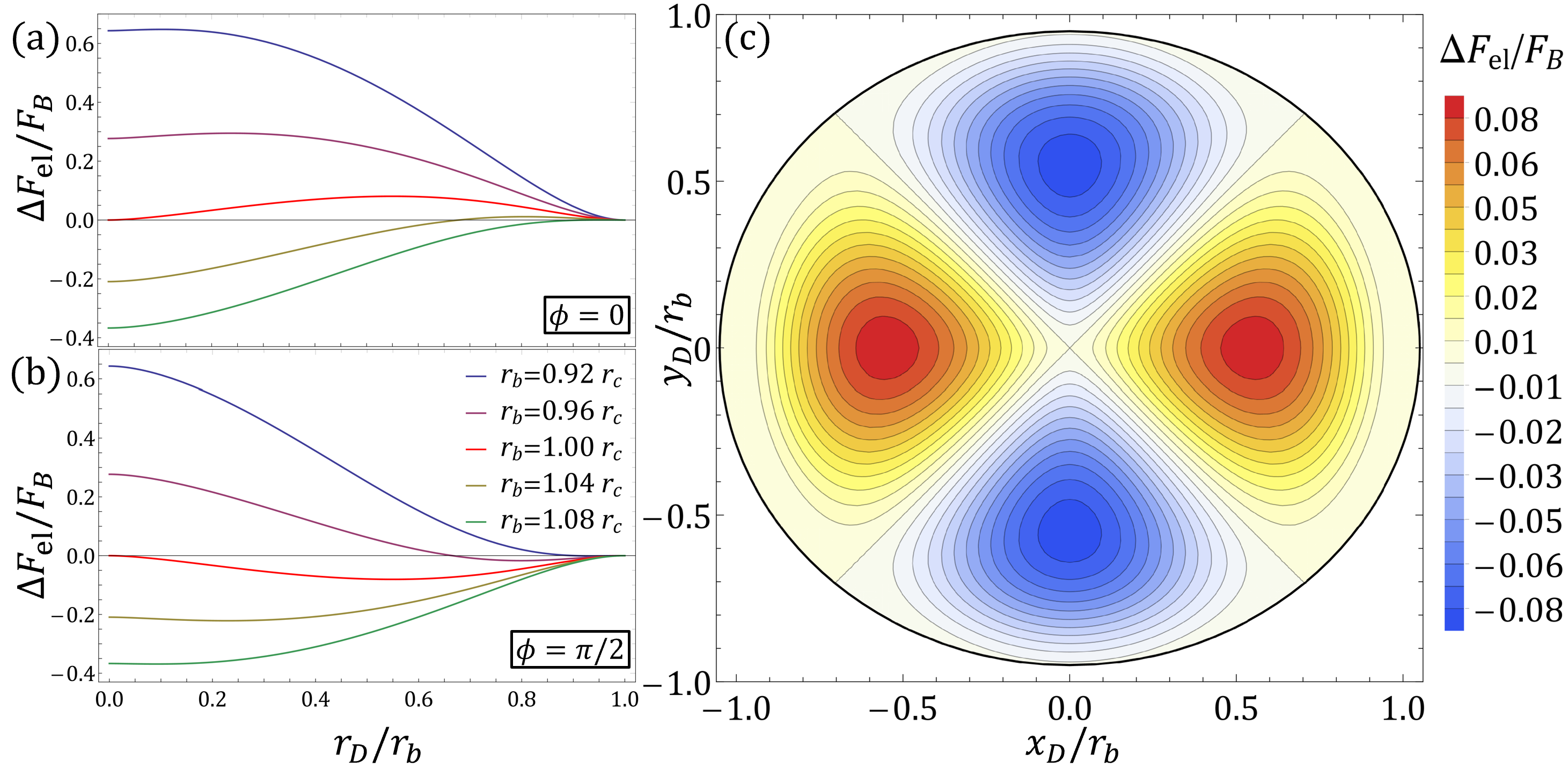}
	\caption{Normalized energy difference $\Delta F_{\text{el}}/F_B$ for an ellipsoid ($\epsilon=0.05$); varying the cap extent $r_b$ around transition, (a) along the major axis ($\phi=0$), an energy maximum persists, whereas there is an energy minimum along (b) the minor axis ($\phi=\pi/2$) -- thus the preferred direction of defect migration is predicted to be along the minor axis. The energy landscape is displayed in (c) showing the location of these global maxima/minima at transition, i.e. at $r_b=r_c$. Note that these plots are independent of $\kappa$ since we replace $r_b$ in \eqref{DelFel_ellipse} by multiples of the small-slope value $r_c=\sqrt{2/3}/\kappa$.}\label{fig6_ellipse}
\end{figure}

We display in Fig.~\ref{fig6_ellipse}, Eq.~\eqref{DelFel_ellipse} as a function of $\overline{r}_D$ for $\epsilon=0.05$ (eccentricity of $e\approx 0.45$) along (a) the major axis ($\phi_D=0$) and (b) the minor axis ($\phi_D=\pi/2$). As expected, breaking the circular cross-sectional symmetry forces the system to select a preferred direction of migration of the defect, which in this situation is the minor axis: this perturbed small-slope formalism predicts a continuous variation of the defect position along $\phi=\pi/2$ until the apex position is established for a certain energy difference (here, $\Delta F_{\text{el}}/F_B\approx -0.35$ as marked by the green curve in Fig.~\ref{fig6_ellipse}(b)). 

To quantify the scale of the secondary structure, we set $\Delta F_{\text{el}}(0)=0$ in $\eqref{DelFel_ellipse}$ to obtain the transition criterion $r_c = \sqrt{2/3}/\kappa$ (unchanged from the rotationally symmetric situation), while the position of the maxima/minima obtained by setting $\Delta F_{\text{el}}'(0)=0$ results in $r_D\approx 0.55r_b$, a slight but significant difference from the position for intrinsic secondary structures described in section~\ref{secanalytica}. Finally, inserting into $\eqref{DelFel_ellipse}$ we obtain the energy scale  
\begin{align}
    \frac{\Delta F^{\text{sec}}_{\text{el}}}{F_B} \approx 1.61 \epsilon \cos 2\phi_D +\mathcal{O}\left(\epsilon^2\right)\,.
    \label{delfellipse}
\end{align}
At the critical cap extent $r_b=r_c$, the small-slope $\Delta F_{\text{el}}$ vanishes, so that the secondary energy structure is proportional to $\epsilon$ and quadrupolar in $\phi_D$. Within the small-slope approximation this correction does not depend on $\kappa$ when normalized by $F_B$. The magnitude of secondary energy maxima and minima (cf.\ Fig.~\ref{fig6_ellipse}) is well described by \eqref{delfellipse}. Having identified a universal scale of {\em intrinsic} secondary structure maxima in section~\ref{secanalytica} as $\Delta F_{\text{el}}^{\text{sec}}/F_B\approx 0.034$ for moderate to large $\kappa$, we can say that symmetry breaking is likely to transform the transition character from discontinuous to continuous along the minor axis beyond a strain of $\epsilon\gg 0.02$. Quite subtle symmetry breaking is thus capable of qualitatively changing the disorder transition. 
We note that inclusion of the next order $\mathcal{O}(r_b^4)$ terms introduces non-trivial couplings between the shape perturbation and the secondary structure discussed in the previous subsections, leading to modification of \eqref{delfellipse}. We know from Sec.~\ref{secanalytica} that the leading term of the transition criterion $r_c$ becomes approximately $a=0.844$. Defining the deviation of $a$ from the small-slope value as $\delta=a-\sqrt{2/3}$, a Taylor expansion of \eqref{DelFel_ellipse} in $\delta\ll 1$ alters the numerical prefactor of \eqref{delfellipse} to $(1.61-7.9\delta)\approx 1.39$. This is a small quantitative deviation that does not change the nature of the effects discussed above. 

\section{Conclusions}\label{secconcl}

The present work demonstrates that simple, general criteria can be derived not just for the onset of energetically favored disclination disorder on curved open surfaces, but to predict the secondary structure of that disorder transition.

The secondary structure of the transition, i.e., whether the displacement of the disclination occurs continuously or discontinuously), is important for predictions of actual defect placement in practical applications. In particular, symmetry-breaking defect positions can be energetically favorable over well-defined ranges of parameters, such as the extent of the open surface. We have shown that these secondary effects occur over a small, quantifiable range of sizes around the onset of disorder. Similarly, the elastic energy of the surface universally changes by a well-defined amount as the defect position changes -- a few percent of the total energy.

Accordingly, this secondary structure can be altered by relatively small modifications of the mechanics of the problem. In particular, even slight anisotropy in the shape of the surface will force the optimal position of the disclination onto one of the principal axes, and will make defect displacements continuous even if they are intrinsically discontinuous. 

It is noteworthy that the analytical approximations yielding results for the secondary structure of the transition require more detailed knowledge of the surface shape than those that allow for an evaluation of the onset of disorder. In particular, the fourth apical derivative of the surface shape (by symmetry the next order after the apical curvature) is a strong determinant of the transition characteristics. Likewise, our formalism makes use of the second derivative of the non-local weighted Gaussian curvature, $\Gamma_S^{\prime\prime}(0)$, as opposed to just its functional value. How accurately these quantities can be determined in an application, and how the presence of positional disorder (dislocations) may alter the results will be the subject of future study.

For a given shape with mobile disclinations, the current work offers easy-to-check criteria for whether the onset of disorder results in  robust central defect placement (discontinuous transition) or whether a variety of configurations may be observed (continuous transition). In applications of shells with Gaussian curvature, be they viral capsids, tissue structures like insect eyes, or optical engineering systems such as microlens arrays, these insights also provide bounds on the degree of symmetry needed to maintain an ordered lattice on such surfaces.

\section*{Conflicts of interest}
There are no conflicts to declare

\section*{Acknowledgements}
The authors are grateful to Mark Bowick and Gregory Grason for insightful discussions, and to the Grason group for making Mathematica code for plotting defects available. SA acknowledges partial support by the NSF under grant $\#$ 1504301.




\renewcommand\refname{References}

\bibliographystyle{apsrev4-2}
\bibliography{ms} 

\end{document}


\title{Electronic Supplementary Information to Predicting the Characteristics of Defect Transitions on Curved Surfaces}
	\author{Siddhansh Agarwal}
	\author{Sascha Hilgenfeldt}%
	\affiliation{Mechanical Science and Engineering, University of Illinois, Urbana-Champaign, Illinois 61801, USA}%
	\maketitle
	

\sloppy 
\preprint{APS/123-QED}
\allowdisplaybreaks

\section{Complete covariant formalism}

We follow the covariant formalism developed by \citet{giomi2007crystalline,bowick2009two} in the following. Let $\mathbb{P}$ be a smooth two-dimensional surface of a crystal lattice in $\mathbb{R}^3$. The elastic free energy of the crystal may be expressed in the form:
\begin{align}
F = F_{\text{el}} +F_c + F_0,
\end{align}
where $F_0$ is the free energy of the defect-free monolayer, $F_c$ is the core energy of defects and $F_{\text{el}}$ is the elastic energy associated with defect interaction. As we will change neither the shape of the surface nor the number of defects, any minimization is governed by $F_{\text{el}}$, which can be written as
\begin{align}
F_{\text{el}} = \frac{Y}{2}\int \mathrm{d} \mathbf{x} \,\mathrm{d} \mathbf{y} \,G_{2L}(\mathbf{x},\mathbf{y}) q_T(\mathbf{x}) q_T(\mathbf{y}),
\end{align}
where $Y$ is the Young's modulus for the planar crystal and $G_{2L}(\mathbf{x},\mathbf{y})$ is the Green's function for the covariant biharmonic operator on the manifold. The quantity $q_T(\mathbf{x})$ represents the effective topological charge density; in the presence of discrete $q=+1$ disclinations at $\mathbf{x}_\alpha$  it takes the form
\begin{align}
q_T(\mathbf{x}) = \sum_\alpha \frac{\pi}{3} \delta(\mathbf{x},\mathbf{x}_\alpha) -K_G(\mathbf{x}),\label{rho}
\end{align}
where $\delta(\mathbf{x},\mathbf{x}_\alpha) = g^{-1/2} \delta (x_1-x_{\alpha_1})\delta (x_2-x_{\alpha_2})$ is the Dirac delta function on the manifold parametrized by $\mathbf{x}=(x_1,x_2)$($=(r,\phi)$ in polar coordinates). The second term $K_G(\mathbf{x})$ is the Gaussian curvature of the surface. On a topological disk with total charge Q=+6, the minimal number of disclinations is 6. Any disclination located at the boundary will not contribute to $F_{\text{el}}$, because $G_{2L}$ vanishes there. In this work, we  compare energies of configurations with all 6 disclinations at the boundary to those with one disclination located at an arbitrary $\mathbf{x}_D$. Therefore, we consider the simpler $q_T(\mathbf{x},\mathbf{x}_D) = \frac{\pi}{3} \delta(\mathbf{x},\mathbf{x}_D) -K_G(\mathbf{x})$.
We begin by observing that the Airy stress function $\chi$ solves the following inhomogeneous biharmonic equation:
\begin{align}
\Delta^2 \chi(\mathbf{x},\mathbf{x}_D) =Y q_T(\mathbf{x},\mathbf{x}_D) \label{biharm},
\end{align}
with no stress boundary conditions
\begin{align}
	\chi(\mathbf{x},\mathbf{x}_D) = 0, \quad \mathbf{x}\in \partial \mathbb{P};\quad	\nu_i \nabla^i \chi(\mathbf{x},\mathbf{x}_D) = 0,\quad \mathbf{x}\in \partial \mathbb{P}. \label{neumann}
\end{align}
The solution of \eqref{biharm} will then be
\begin{align}
\chi(\mathbf{x},\mathbf{x}_D) = \int d^2 y G_L (\mathbf{x},\mathbf{y}) \Gamma(\mathbf{y},\mathbf{y}_D),
\end{align}
where $G_L (\mathbf{x},\mathbf{y})$ is the Green's function of the covariant Laplace operator on $\mathbb{P}$ with Dirichlet boundary conditions
\begin{align}
	\Delta G_L(\mathbf{x},\mathbf{.}) = \delta(\mathbf{x},\mathbf{.}), \quad \mathbf{x}\in \mathbb{P};\quad G_L(\mathbf{x},\mathbf{.}) = 0,\quad \mathbf{x}\in \partial \mathbb{P},\label{Gl}
\end{align}
and $\Gamma(\mathbf{x},\mathbf{x}_D)=\Delta\chi(\mathbf{x},\mathbf{x}_D)$ is the solution of the Poisson problem:
\begin{align}
\Delta\Gamma(\mathbf{x},\mathbf{x}_D) = Y q_T(\mathbf{x},\mathbf{x}_D),
\end{align}
which can be expressed formally as:
\begin{align}
\Gamma(\mathbf{x},\mathbf{x}_D) = Y\int q_T(\mathbf{y},\mathbf{y}_D)G_L(\mathbf{x},\mathbf{y}) d^2 y = -\Gamma_D(\mathbf{x},\mathbf{x}_D) -\Gamma_S(\mathbf{x}) + U(\mathbf{x},\mathbf{x}_D),\label{gammaxapp}
\end{align}
where
\begin{align}
    \Gamma_D(\mathbf{x},\mathbf{x}_D) = - \frac{Y\pi}{3} G_L(\mathbf{x},\mathbf{x}_D),\quad	\Gamma_S(\mathbf{x}) = Y\int K_G(\mathbf{y}) G_L(\mathbf{x},\mathbf{y}) d^2 y, \label{gammadapp}
\end{align}
and $U(\mathbf{x},\mathbf{x}_D)$ is a harmonic function on $\mathbb{P}$ that enforces the Neumann boundary conditions. The first term of \eqref{gammaxapp} represents the bare contribution of disclinations while the second term captures the screening effect of Gaussian curvature. In this paper we restrict ourselves to allowing only one disclination to migrate from the boundary to the apex of the manifold. The Green's function satisfying \eqref{Gl} can be computed explicitly by conformally mapping the surface $\mathbb{P}$ onto the unit disk of the complex plane where the Green's function is known:
\begin{align}
G_L (\mathbf{x},\mathbf{y}) = \frac{1}{2\pi} \log \left|\frac{z(\mathbf{x})-z(\mathbf{y})}{1-z(\mathbf{x}) \overline{z(\mathbf{y})}}\right|,\label{Glapp}
\end{align}
where $z(\mathbf{x}) = \varrho e^{i\phi}$, a point in the unit disk, is the image of a point on the surface $\mathbb{P}$ under the conformal mapping. The Green's function vanishes when the disclination is located at the boundary. For a surface $X(r,\phi)$ with first fundamental form $E=\partial X/\partial r \cdot \partial X/\partial r$, $F=\partial X/\partial r \cdot \partial X/\partial \phi$ and $G=\partial X/\partial \phi \cdot \partial X/\partial \phi$, the metric of the surface will be
\begin{align}
ds^2 = E dr^2 + 2F drd\phi +G d\phi^2 \label{metricX}
\end{align}
whereas the unit disk has the metric
\begin{align}
ds^2 = w(z)\left(\varrho^2 dr^2 + \varrho^2 d\phi^2\right) \label{metricdisk}
\end{align}
where $w(z)$ is a positive conformal weight. The remaining task is now to find the conformal factor $w(z)$ and the conformal radius $\varrho(r)$ by equating the two metrics; these can be explicitly obtained for many rotationally symmetric surfaces but in general, may not be analytically computable.

Taking the two image points on the unit disk as $z(r,\phi)=\varrho_x(r)e^{i\phi}$ and $\zeta(r',\phi')=\varrho_y(r')e^{i\phi'}$, the contribution due to the background Gaussian curvature is split into two parts $\Gamma_S(\mathbf{x}) = \Gamma_{S,1}(\mathbf{x}) - \Gamma_{S,2}(\mathbf{x})$, where
\begin{subequations}
	\begin{align}
	\Gamma_{S,1}(r,\phi) &= \frac{Y}{2\pi} \int d \phi' \, dr' \, \sqrt{g} K(r')\log|z-\zeta|,\\
	\Gamma_{S,2}(r,\phi) &= \frac{Y}{2\pi} \int d \phi' \, dr' \, \sqrt{g} K(r')\log|1-z\overline{\zeta}|,
	\end{align}\label{gammas12}\noindent
\end{subequations}
are evaluated analytically for the specific surfaces considered in this paper. The computation of the harmonic function on the manifold $\mathbb{P}$ is more involved and its contribution to the energy density is given by, 
\begin{align}
U(\mathbf{x},\mathbf{x}_D) = - Y\int d^2 y H(\mathbf{x},\mathbf{y})q_T(\mathbf{y},\mathbf{y}_D),
\end{align}
where $H(\mathbf{x},\mathbf{y})$ is harmonic kernel associated with the Green's function of the weighted biharmonic operator arising from the conformal mapping of the manifold $\mathbb{P}$ onto the unit disk in the complex plane. The harmonic kernel can be written in integral form as \cite{shimorin1997green}
\begin{subequations}
	\begin{align}
	H(z,\zeta)&= -\int_{|\zeta|}^{1}\frac{dt}{\pi t}\int_{0}^{t} ds \, \sqrt{g} k\left[\frac{\varrho^2(s)}{t^2}\zeta \overline{z}\right],\\
	k(z\overline{\zeta}) &= \sum_{n\geq 0} \frac{(z\overline{\zeta})^n}{c_n(1)} + \sum_{n<0} \frac{(\overline{z}\zeta)^{|n|}}{c_{|n|}(1)},\\
	c_n(t)& = 2 \int_{0}^{t} ds \, \sqrt{g} \varrho^{2n}(s).\label{cn}
	\end{align}
\end{subequations}
After making appropriate substitutions, one obtains
\begin{align}
H(z,\zeta) &=-\int_{|\zeta|}^{1}\frac{dt}{\pi t}\int_{0}^{t} ds \, \sqrt{g} \left[\frac{1}{c_0(1)} + \sum_{n\geq 1}\frac{1}{c_n(1)}\frac{\varrho^{2n}(s)}{t^{2n}}\varrho_x^n \varrho_y^n \left(e^{in(\phi-\phi')}+e^{-in(\phi-\phi')}\right)
\right]\nonumber\\
&= -\int_{|\zeta|}^{1}\frac{dt}{\pi t}\int_{0}^{t} ds \, \sqrt{g} \left[\frac{1}{c_0(1)} + \sum_{n\geq 1}\frac{2}{c_n(1)}\frac{\varrho^{2n}(s)}{t^{2n}}\varrho_x^n \varrho_y^n \cos n(\phi-\phi')\right]\nonumber\\
& = -\int_{\varrho_y}^{1}\frac{dt}{\pi t}\left[\frac{1}{2}\frac{c_0(t)}{c_0(1)} + \sum_{n\geq 1}\frac{c_n(t)}{c_n(1)} \frac{2}{t^{2n}}\varrho_x^n \varrho_y^n \cos n(\phi-\phi') \right]\nonumber\\
& = -\frac{1}{2\pi} f_0(\varrho_y) - \sum_{n\geq 1} \frac{1}{\pi}\varrho_x^n \varrho_y^n \cos n(\phi-\phi') f_n(\varrho_y),\noindent
\end{align}
where 
\begin{align}
f_n(\varrho_y)=\int_{\varrho_y}^{1}\frac{c_n(t)}{c_n(1)t^{2n+1}}dt.\label{fn}
\end{align}
Note that the variable $s$ lies on the manifold $\mathbb{P}$, whereas $t$ lies on the unit disk in $\mathbb{R}^2$. Therefore, explicitly
\begin{align}
U(r,\phi,r_D,\phi_D) = -Y\int_{0}^{2\pi} d\phi' \int_{0}^{r_b} & dr' \, \sqrt{g}(r')\,q_T(r',\phi',r_D,\phi_D) \nonumber\\
&\left[ \left(-\frac{1}{2\pi} f_0(\varrho_y) - \sum_{n\geq 1} \frac{1}{\pi}\varrho_x^n \varrho_y^n \cos n(\phi-\phi') f_n(\varrho_y)\right)\right].\label{Uapp}
\end{align}
Here, $f_0$ is the azimuthally symmetric contribution while the higher order modes capture asymmetries arising due to intermediate singularity positions. $\varrho$ is the effective topological charge density as defined according to \eqref{rho} and $Y$ is the material Young's modulus. We note that the isotropic contribution, i.e., due to $f_0$ can be shown to be the same as reported in our previous work \cite{agarwal2020} and is equivalent to $U = \frac{1}{A}\int \Gamma dA$ for rotationally symmetric surfaces with symmetric defect placement.

\section{Breaking rotational symmetry}

As stated in the main text, we apply a shape perturbation to a general surface of revolution, generating an ellipse from a circular boundary by stretching/contracting perpendicular axes by $\epsilon$. Thus,
\begin{align}
x=(1+\epsilon)r\cos \phi,\quad y=(1-\epsilon)r \sin \phi, \quad z= f(r)
\end{align}
with $\epsilon\ll 1$. The metric tensor in the small-slope limit is 
\begin{align}
g_{ij}=\begin{bmatrix} 
	1 + \epsilon(2+\epsilon)\cos 2\phi & -2r\epsilon \sin 2\phi \\
	-2r\epsilon \sin 2\phi & r^2\left(1 + \epsilon(\epsilon-2)\cos 2\phi\right)
\end{bmatrix},
\end{align}
where $\sqrt{g}=r(1-\epsilon^2)$ and eccentricity $e=\sqrt{1-\frac{(1-\epsilon)^2}{(1+\epsilon)^2}}=2\sqrt{\epsilon} + \mathcal{O}(\epsilon^{3/2})$, while the Gaussian curvature is constant to $\mathcal{O}(\epsilon)$ and is given by 
\begin{align}
K_G(r,\phi) = \kappa^2 +\mathcal{O}\left(\epsilon^2\right),
\end{align}
where $\kappa=f''(0)$ is the apical curvature. The elliptical boundary of a section cut parallel to the $xy$-plane is given by
\begin{align}
r(\phi) = \frac{r_b(1-\epsilon^2)}{\sqrt{1+\epsilon(\epsilon-2)\cos 2\phi}} \approx r_b(1+\epsilon \cos 2\phi).
\label{ellipse}
\end{align}
Using a regular perturbation expansion for $\chi =\chi_0 + \epsilon \chi_1$ (and consequently the stresses $\sigma$) while also expanding the Laplace-Beltrami operator for small $\epsilon$ ($\nabla_\perp^2 \approx \nabla_0^2 +\epsilon\nabla_1^2$), we solve the following biharmonic equation for $\chi$ order-wise up to $O(\epsilon)$:
\begin{align}
\frac{1}{Y}(\nabla_0^2 +\epsilon\nabla_1^2)(\nabla_0^2 +\epsilon\nabla_1^2) (\chi_0 + \epsilon \chi_1) &=  \frac{\pi}{3}\frac{\delta(r-r_D)\delta(\phi-\phi_D)}{\sqrt{g}}-K_G(r,\phi) \label{biharmperp}.
\end{align}
The Airy stress $\chi$ relates to the stress components via the usual relations in polar coordinates:
\begin{align}
\sigma_{rr} = \frac{1}{r}\frac{\partial \chi}{\partial r} + \frac{1}{r^2} \frac{\partial^2 \chi}{\partial \phi^2},\quad \sigma_{r\phi}= -\frac{\partial}{\partial r}\left(\frac{1}{r} \frac{\partial \chi}{\partial \phi}\right),\quad \sigma_{\phi\phi} = \frac{\partial^2 \chi}{\partial r^2}.
\end{align}
For $\epsilon\ll 1$, the normal vector components can be expanded as: $n_r = 1+ O(\epsilon^2)$ and $n_\phi = 2\epsilon \sin 2\phi + O(\epsilon^2)$. Consistently expanding both the stress components and their arguments ($r(\phi)=r_b + \epsilon r_b\cos 2\phi$), one obtains up to $O(\epsilon)$:
\begin{subequations}
\begin{align}
\sigma_{rr_0}(r_b) + \epsilon\left( r_b \cos 2\phi \frac{\partial \sigma_{rr_0}}{\partial r}\bigg|_{r_b} +2\sigma_{r\phi_0}(r_b)\sin 2\phi +\sigma_{rr_{1}}(r_b)\right) &= 0 \\
\sigma_{r\phi_0}(r_b)+\epsilon\left(r_b \cos 2\phi \frac{\partial \sigma_{r\phi_0}}{\partial r}\bigg|_{r_b}+2\sigma_{\phi\phi_0}(r_b) \sin 2\phi + \sigma_{r\phi_1}(r_b)\right) &=0 
\end{align}
\end{subequations}
For a stress-free boundary the elastic energy is written simply in terms of the trace of the stress tensor \cite{giomi2007crystalline}, i.e., $\Gamma(r,\phi)\approx\Gamma_0(r,\phi)+\epsilon\Gamma_1(r,\phi)=\sigma_{rr_0} + \sigma_{\phi \phi_0}+\epsilon(\sigma_{rr_1} + \sigma_{\phi \phi_1})$. Therefore, the elastic energy formally reads:
\begin{align}
F_{\text{el}} &= \frac{1}{2Y}\int_0^{2\pi}\int_0^{r_b} \sqrt{g} \Gamma(r,\phi)^2 dr d\phi\nonumber\\
&=  \frac{1}{2Y}\int_0^{2\pi}\int_0^{r_b} r \left(\Gamma_0(r,\phi)^2 +2 \epsilon\Gamma_0(r,\phi)\Gamma_1(r,\phi)\right)dr d\phi +\mathcal{O}(\epsilon^2),
\end{align}
which is the expression given in the main text. This energy integral is executed analytically exploiting the orthogonality of trigonometric functions.

While \citet{azadi2016neutral} focused on the azimuthally symmetric center or boundary placement of the defect, following work going back to \citet{michell1899direct}, we use a Fourier series expansion of the Airy stress function $\chi(r,\phi)=a_0(r)+ \sum_{n=1}^\infty \left(a_n(r)\cos n\phi +  b_n(r)\sin n\phi\right)$, to solve this system of equations for a general position $(r_D,\phi_D)$ of the disclination. The stress components expressed in terms of this expansion read
\begin{subequations}
	\begin{align}
\sigma_{rr} &= \frac{1}{r}\frac{\partial \chi}{\partial r} + \frac{1}{r^2} \frac{\partial^2 \chi}{\partial \phi^2}\nonumber\\
&=\frac{1}{r}\left[a_0^{(1)}(r) + \sum_{n=1}^{\infty}\left\{\left(a_n^{(1)}(r)-n^2\frac{a_n(r)}{r}\right)\cos n \phi + \left(b_n^{(1)}(r)-n^2\frac{b_n(r)}{r}\right)\sin n \phi\right\}\right],\\
\sigma_{r\phi}&= -\frac{\partial}{\partial r}\left(\frac{1}{r} \frac{\partial \chi}{\partial \phi}\right)=\frac{1}{r}\sum_{n=1}^{\infty}n\left[\left( a_n^{(1)}(r)-\frac{a_n(r)}{r}\right)\sin n\phi-\left( b_n^{(1)}(r)-\frac{b_n(r)}{r}\right)\cos n\phi\right],\\
\sigma_{\phi\phi}& = \frac{\partial^2 \chi}{\partial r^2}=a_0^{(2)}(r) + \sum_{n=1}^{\infty}\left(a_n^{(2)}(r)\cos n\phi + b_n^{(2)}(r)\sin n\phi\right).
\end{align}
\end{subequations}
\subsection{Zeroth order solution}

At $O(1)$, we employ a Fourier series expansion of the Airy stress function such that $\chi_0(r,\phi) = a_0(r)+\sum_{n=1}^{\infty}a_n(r)\cos n(\phi-\phi_D)$. The solution will be azimuthally symmetric only when the singularity is placed at the center or at the boundary of the surface \cite{azadi2016neutral}. This results in:
\begin{align}
\frac{1}{Y}\nabla^4_0 \chi_{0}(r,\phi) = -\kappa^2+\frac{\pi}{3}\frac{\delta(r-r_D)}{\pi r}\left[\frac{1}{2}+ \sum_{n=1}^{\infty}\cos(n(\phi-\phi_D))\right] \label{biharmpde0}
\end{align}
where $\nabla^2_0 =  \frac{\partial^2}{\partial r^2} + \frac{1}{r}\frac{\partial}{\partial r} +\frac{1}{r^2}\frac{\partial^2}{\partial \phi^2}$ and subject to the radial stress component boundary condition ($\sigma_{rr_0}(r_b)=0$, $\sigma_{r\phi_0}(r_b)=0$). The Fourier expansion of the stress components in terms of these modes is as follows:
\begin{subequations}
\begin{align}
\sigma_{rr_0} &= \frac{1}{r}\frac{\partial \chi}{\partial r} + \frac{1}{r^2} \frac{\partial^2 \chi}{\partial \phi^2}=\frac{1}{r}\left[a_0^{(1)}(r) + \sum_{n=1}^{\infty}\left(a_n^{(1)}(r)-n^2\frac{a_n(r)}{r}\right)\cos n (\phi-\phi_D)\right],\\
\sigma_{r\phi_0}&= -\frac{\partial}{\partial r}\left(\frac{1}{r} \frac{\partial \chi}{\partial \phi}\right)=\frac{1}{r}\sum_{n=1}^{\infty}n\left( a_n^{(1)}(r)-\frac{a_n(r)}{r}\right)\sin n (\phi-\phi_D),\\
\sigma_{\phi\phi_0}& = \frac{\partial^2 \chi}{\partial r^2}=a_0^{(2)}(r) + \sum_{n=1}^{\infty}a_n^{(2)}(r)\cos n (\phi-\phi_D).
\end{align}
\end{subequations}
The system of ODEs one thus obtains is:
\begin{subequations}
\begin{align}
\frac{1}{Y}\left[a_0^{(4)}(r)+\frac{2}{r}a_0^{(3)}(r)-\frac{1}{r^2}a_0^{(2)}(r)+\frac{1}{r^3}a_0^{(1)}(r)\right]&= -\kappa^2+ \frac{\delta(r-r_D)}{6 r}\\
a_n^{(4)}(r)+\frac{2}{r}a_n^{(3)}(r)-\frac{(1+2n^2)}{r^2}a_n^{(2)}(r)+\frac{1+2n^2}{r^3}a_n^{(1)}(r)+\frac{n^2(n^2-4)}{r^4}a_n(r)&= \frac{Y\delta(r-r_D)}{3 r},
\end{align}
\end{subequations}
where $^{()}$ indicates differentiation with respect to $r$, along with the following system of boundary conditions
\begin{subequations}
\begin{align}
\frac{1}{r_b}\left[a_0^{(1)}(r_b) + \sum_{n=1}^{\infty}\left(a_n^{(1)}(r_b)-n^2\frac{a_n(r_b)}{r_b}\right)\cos n(\phi-\phi_D)  \right] &= 0 \\
\frac{1}{r_b}\left[\sum_{n=1}^{\infty}n\left( a_n^{(1)}(r_b)-\frac{a_n(r_b)}{r_b}\right)\sin n(\phi-\phi_D) \right]&=0
\end{align}
\end{subequations}
The leading order (small-slope) Fourier amplitudes for a rotationally symmetric surface with central curvature $\kappa$ in the presence of an arbitrarily positioned disclination defect is given by:
\begin{align}
a_0(r) = &-\frac{Y\kappa^2r^2}{64}(r^2-2r_b^2) + \frac{Y}{24}H(r-r_D)\left(r_D^2 -r^2 +(r_D^2+r^2)\log\left(\frac{r}{r_D}\right) \right) \nonumber\\
&+ \frac{Y}{48}r^2\left(1-\frac{r_D^2}{r_b^2}+2\log\left(\frac{r_D}{r_b}\right)\right),\\
a_n(r)=& \frac{Y}{24}\frac{H(r-r_D)}{ n(n^2-1)}\left[\left(\frac{r_D}{r}\right)^n\left((n+1)r^2-(n-1)r_D^2\right)+\left(\frac{r}{r_D}\right)^n\left((n-1)r^2-(n+1)r_D^2\right)\right]\nonumber\\
-&  \frac{Y}{24}\frac{\left(r/r_D\right)^n}{ n(n^2-1)}\bigg[(n-1)r^2-(n+1)r_D^2 \nonumber\\
&+ \left(\frac{r_D}{r_b}\right)^{2n}\left(n(n-1)\left(\frac{rr_D}{r_b}\right)^{2}-(n^2-1)(r^2+r_D^2)+n(n+1)r_b^2\right)\bigg],
\end{align}
where $H$ is the Heaviside function and $a_0(r)$ is the azimuthally symmetric solution reported in \citep{azadi2016neutral}. All the $n\neq 0$ modes go to zero when the singularity is decorated at the boundary or at the center of the surface but not otherwise.

\subsection{First order solution}

At $O(\epsilon)$, analogous to the zeroth order but more generally, we employ a Fourier series decomposition of the Airy stress function such that, $\chi_1(r,\phi) = c_0(r) +\sum_{n=1}^{\infty}\left(c_n(r)\cos n\phi+d_n(r)\sin n\phi \right)$. This results in
\begin{align}
\frac{1}{Y}\left[\nabla^4_0 \chi_{1}(r,\phi) +2\nabla^2_1(\nabla^2_0 \chi_{0}(r,\phi)) \right]=0 \label{biharmpde1}
\end{align}
where $\nabla^2_1 = 2\cos^2\phi \frac{\partial^2}{\partial r^2} + 2\sin \phi \left(\frac{2\cos\phi}{r^2}\frac{\partial}{\partial \phi}+\frac{\sin\phi}{r}\frac{\partial}{\partial r}\right) + 2\sin \phi \left(\frac{\sin\phi}{r^2}\frac{\partial^2}{\partial \phi^2}-\frac{2\cos\phi}{r}\frac{\partial^2}{\partial r \partial \phi}\right)$ and subject to the following boundary conditions:
\begin{subequations}
\begin{align}
r_b \cos 2\phi \frac{\partial \sigma_{rr_0}}{\partial r}\bigg|_{r_b} + \cancelto{0}{2\sigma_{r\phi_0}(r_b)}\sin 2\phi+ \sigma_{rr_{1}}(r_b)&=0,\\
r_b \cos 2\phi \frac{\partial \sigma_{r\phi_0}}{\partial r}\bigg|_{r_b}+2\sigma_{\phi\phi_0}(r_b) \sin 2\phi + \sigma_{r\phi_1}(r_b) &=0.
\end{align}\label{Oeps_BC}\noindent
\end{subequations}

Inserting the Fourier expansion, \eqref{biharmpde1} reduces to the following system of ODEs that now have a forcing term from the previous order:
\begin{subequations}
\begin{align}
&c_0^{(4)}(r)+\frac{2}{r}c_0^{(3)}(r)-\frac{1}{r^2}c_0^{(2)}(r)+\frac{1}{r^3}c_0^{(1)}(r)=-2 \left(\frac{2 r^2 a_0{}^{(3)}(r)-r a_0^2(r)+a_0^1(r)}{r^3}+a_0{}^{(4)}(r)\right) \nonumber\\
&-\frac{\cos \left(2 \phi _D\right) \left(3 a_2'(r)+r \left(r \left(4 a_2{}^{(3)}(r)+r a_2{}^{(4)}(r)\right)-3 a_2''(r)\right)\right)}{r^3}\\
&c_n^{(4)}(r)+\frac{2}{r}c_n^{(3)}(r)-\frac{(1+2n^2)}{r^2}c_n^{(2)}(r)+\frac{1+2n^2}{r^3}c_n^{(1)}(r)+\frac{n^2(n^2-4)}{r^4}c_n(r)= \nonumber\\
&-\frac{2 \cos \left(n \phi _D\right)}{r^4} \bigg(r \left(\left(2 n^2+1\right) a_n'(r)+r \left(\left(-2 n^2-1\right) a_n''(r)+r \left(2 a_n{}^{(3)}(r)+r a_n{}^{(4)}(r)\right)\right)\right)\nonumber\\
&+\left(n^2-4\right) n^2 a_n(r)\bigg)-\frac{\cos \left((n-2) \phi _D\right)}{r^4} \bigg(r \bigg((2 (n-2) (n-1) n+3) a_{n-2}'(r)\nonumber\\
&+r \left(r \left(r a_{n-2}{}^{(4)}(r)-2 (n-2) a_{n-2}{}^{(3)}(r)\right)-3 a_{n-2}''(r)\right)\bigg)-(n-2)^2 n (n+2) a_{n-2}(r)\bigg)\nonumber\\
&-\frac{\cos \left((n+2) \phi _D\right)}{r^4} \bigg(r \bigg((3-2 n (n+1) (n+2)) a_{n+2}'(r)\nonumber\\
&+r \left(r \left(2 (n+2) a_{n+2}{}^{(3)}(r)+r a_{n+2}{}^{(4)}(r)\right)-3 a_{n+2}''(r)\right)\bigg)+(2-n) n (n+2)^2 a_{n+2}(r)\bigg)\\
&d_n^{(4)}(r)+\frac{2}{r}d_n^{(3)}(r)-\frac{(1+2n^2)}{r^2}d_n^{(2)}(r)+\frac{1+2n^2}{r^3}d_n^{(1)}(r)+\frac{n^2(n^2-4)}{r^4}d_n(r)= \nonumber\\
&-\frac{2 \sin \left(n \phi _D\right)}{r^4} \bigg(r \left(\left(2 n^2+1\right) a_n'(r)+r \left(\left(-2 n^2-1\right) a_n''(r)+r \left(2 a_n{}^{(3)}(r)+r a_n{}^{(4)}(r)\right)\right)\right)\nonumber\\
&+\left(n^2-4\right) n^2 a_n(r)\bigg)-\frac{\sin \left((n-2) \phi _D\right)}{r^4} \bigg(r \bigg((2 (n-2) (n-1) n+3) a_{n-2}'(r)\nonumber\\
&+r \left(r \left(r a_{n-2}{}^{(4)}(r)-2 (n-2) a_{n-2}{}^{(3)}(r)\right)-3 a_{n-2}''(r)\right)\bigg)-(n-2)^2 n (n+2) a_{n-2}(r)\bigg)\nonumber\\
&-\frac{\sin \left((n+2) \phi _D\right)}{r^4} \bigg(r \bigg((3-2 n (n+1) (n+2)) a_{n+2}'(r)\nonumber\\
&+r \left(r \left(2 (n+2) a_{n+2}{}^{(3)}(r)+r a_{n+2}{}^{(4)}(r)\right)-3 a_{n+2}''(r)\right)\bigg)+(2-n) n (n+2)^2 a_{n+2}(r)\bigg)
\end{align}
\end{subequations}
where $n=1,2,3,4..$. These together with \eqref{Oeps_BC} are solved analytically to obtain the $\mathcal{O}(\epsilon)$ Fourier coefficients whose lengthy expressions are available upon request.

\bibliographystyle{apsrev4-2}
\bibliography{supplement} 